\documentclass[a4paper,11pt]{article}
\usepackage{jheppub} 
\usepackage{amsmath, amssymb, amsthm, epsfig, fancyhdr,epsfig,multirow}
\usepackage[utf8]{inputenc}
\usepackage{amsmath}
\usepackage{amsfonts}
\usepackage{url}
\usepackage{amssymb}
\usepackage{xcolor}
\usepackage{comment}
\usepackage{subcaption}
\usepackage[normalem]{ulem}
\usepackage{tabularx}
\usepackage{comment}
\usepackage{float}
\usepackage{graphicx}
\usepackage{appendix}
\usepackage[most]{tcolorbox}
\usepackage{physics}
\usepackage{mathtools}
\usepackage[normalem]{ulem}

\usepackage{tikz}
\usepackage{tikz-feynman}
\usepackage{braket}
\usepackage{cancel}
\usepackage{slashed}
\usetikzlibrary{decorations.pathmorphing}

\title{Positivity in Massive Spin-3/2 EFTs and the Planck-Suppressed  Neighbourhood of Supergravity}

\author[a,1]{Jay Desai}
\author[a,2]{Diptimoy Ghosh}
\author[b,3]{Saurabh Pant}

\affiliation[a]{Department of Physics, Indian Institute of Science, Education and Research, Pune, India}
\affiliation[b]{Department of Physics and Center for Field Theory and Particle Physics, Fudan University,
Shanghai 200438, China}

\emailAdd{desai.jay@students.iiserpune.ac.in}
\emailAdd{diptimoy.ghosh@iiserpune.ac.in}
\emailAdd{saurabhpant@fudan.edu.cn}

\abstract{It is well known that a strictly massless spin-$3/2$ particle can interact consistently only within supergravity. Recently, positivity arguments have shown that an effective field theory of a massive Majorana spin-$3/2$ particle admits  an arbitrarily small mass only if a graviton is present and the four-fermion contact interactions are tuned to the values dictated by $\mathcal{N}=1$ supergravity. In this work, we investigate how this limit is approached at finite mass. Assuming that the graviton $t$-channel pole can be discarded, we derive non-forward, tree-level dispersive bounds on massive spin-3/2 contact operators and determine the region of effective couplings consistent with unitarity and analyticity. For sufficiently small $m$, we find that the allowed parameter space forms a bounded, Planck-suppressed neighbourhood of the supergravity point, defined by the supergravity values of the four-fermion couplings. The supergravity point lies on the boundary of this region. In the regime $m \ll M_{\rm Pl}$, the volume of the allowed region scales parametrically as
\[
\mathrm{Vol} \sim \frac{m^{6}}{M_{\rm Pl}^{6}} \, ,
\]
and shrinks to zero as $m \to 0$, smoothly reproducing the massless-limit results. The allowed region becomes unbounded when mass approaches the Planck scale. We further analyze the effect of including additional light scalar and pseudo-scalar degrees of freedom, motivated by the Polonyi model, and find that their couplings are also bounded in a way similar to the contact couplings and that it doesn't enlarge the allowed contact coupling space.}

\begin{document}
\maketitle
\flushbottom
\section{Introduction}

The interactions of particles with spin greater than one are known to be severely constrained by consistency requirements such as Lorentz invariance, locality, and unitarity. In particular, it has long been understood that a strictly massless spin-$3/2$ particle cannot interact consistently in flat spacetime unless it is coupled to gravity in a supersymmetric manner. This conclusion has been established from several complementary perspectives, including soft-theorem arguments \cite{GRISARU1977323} and modern on-shell amplitude methods \cite{McGady:2013sga}.\\\\
More recently, these ideas have been extended to the case of a \emph{massive} spin-$3/2$ particle. Using on-shell constructibility and assuming a smooth massless limit, Refs.~\cite{Gherghetta:2024tob,Gherghetta:2025tlx} showed that the four-point gravitino contact interactions are uniquely fixed to those of $\mathcal{N}=1$ supergravity (SUGRA). We review this analysis in Appendix \ref{on_shell_summary}. By further requiring that the partial wave unitarity cutoff of the effective field theory (EFT) remain independent of $M_{\rm Pl}$, they recovered the structure of the Polonyi model. Complementary constraints were obtained in \cite{Bellazzini:2025shd} using analyticity and positivity arguments (see \cite{Adams:2006sv, Vecchi:2007na,Manohar:2008tc,Low:2009di,Nicolis:2009qm, Bellazzini:2014waa,Bellazzini:2016xrt,Caron-Huot:2016cwu, deRham:2017avq, deRham:2017zjm,Bellazzini:2017fep, Zhang:2018shp, Bellazzini:2020cot, Tolley:2020gtv, Caron-Huot:2020cmc,Hebbar:2020ukp,Sinha:2020win,Arkani-Hamed:2020blm, Caron-Huot:2021rmr,Henriksson:2021ymi,Arkani-Hamed:2021ajd,Azatov:2021ygj,Bellazzini:2021oaj, Creminelli:2022onn, EliasMiro:2022xaa,Ghosh:2022qqq,Haring:2022sdp,Li:2022aby,Mizera:2023tfe,Berman:2023jys,Hui:2023pxc,Creminelli:2023kze,Bertucci:2024qzt,Creminelli:2024lhd, Caron-Huot:2024tsk,Beadle:2024hqg,Chakraborty:2024ciu, Peng:2025klv, Beadle:2025cdx, Desai:2025alt,Liao:2025npz, Ye:2025zhs,Liu:2025deo} for earlier work). In particular, it was shown that an EFT of a massive Majorana spin-$3/2$ particle $\psi_\mu$ admits an arbitrarily small mass only if a graviton is present and the $\psi_\mu^4$ contact interactions are tuned precisely to their supergravity values.\\\\
Taken together, these results establish that supergravity emerges as the unique infrared-consistent endpoint of the theory at leading order in the expansion around $m\ll M_{\rm Pl}$. However, while the coupling space in the regime $m\ll M_{\rm Pl}$ is now well understood to be at the supergravity point, the effects of finite mass sub-leading corrections are not known. In particular, once the spin-3/2 mass is kept non-zero but small compared to $M_{\rm Pl}$, it is natural to ask whether analyticity and unitarity permit a finite neighbourhood of couplings around the supergravity point, or whether supergravity remains isolated even at finite $m$.\\\\
The purpose of this paper is to explore finite-mass effects around the supergravity point in the space of allowed contact couplings in greater detail, using the non-forward dispersive framework of \cite{Bellazzini:2025shd} (valid at finite mass), which previously established the supergravity point as the leading-order result.
We derive tree-level, non-forward positivity bounds for EFTs of massive, Majorana spin-$3/2$ particles, keeping the mass arbitrary but parametrically below the Planck scale $M_{\rm Pl}$. Unlike forward-limit bounds, non-forward dispersive constraints probe a broader set of interactions and provide stronger control over longitudinal $2 \to 2$ scattering amplitudes.\\\\
We begin in section~\ref{Only Contact} by analyzing theories containing only dimension-six contact interactions of the spin-$3/2$ field. In this case, analyticity enforces a lower bound on $m/\Lambda$, excluding an arbitrarily small mass regime and demonstrating that contact interactions alone cannot support a smooth approach to $m \to 0$. \\\\
We then turn in section~\ref{Neighbourhood of Supergravity} to a supergravity-motivated setup in which a graviton is minimally coupled to $\psi_\mu$, while allowing for deviations from the exact supergravity contact couplings.  (There is of course obstruction to dispersive arguments from the $t$-channel pole in the forward limit from graviton exchange diagrams, which is addressed in section \ref{t-pole}.) In this case, the structure of the allowed parameter space changes qualitatively. For sufficiently small $m$, the dispersive bounds admit a bounded region in coupling space. In the regime $m \ll M_{\rm Pl}$, the volume of this region scales parametrically as
\begin{equation}
\mathrm{Vol} \sim \frac{m^6}{M_{\rm Pl}^6},
\end{equation}
and shrinks to zero as $m \to 0$, with the supergravity point lying on the boundary of the allowed region. As $m$ approaches the Planck scale, the allowed region undergoes a qualitative transition and unbounded directions in coupling space can emerge.\\\\
In section~\ref{Light Scalars}, we investigate whether the inclusion of additional light scalar and pseudo-scalar degrees of freedom, motivated by the Polonyi model, can enlarge the dispersively allowed region. While such fields can soften the high-energy growth of longitudinal amplitudes and raise the perturbative unitarity cutoff, we find that they do not expand the Planck-suppressed neighbourhood around the supergravity point in the small-mass regime.\\\\
Finally, in section \ref{Partial Wave Unitarity}, we also derive complementary bounds imposing partial wave unitarity of the amplitudes up to a scale within the EFT, and compare the qualitative nature of these bounds with those obtained from dispersive analysis.\\ \\ 
Our results therefore provide a quantitative characterization of the finite-mass neighbourhood of supergravity selected by analyticity and unitarity, clarifying how the strict $m \to 0$ consistency constraints are approached at finite mass.

\section{Set up}
In this section, we first set up the notations, conventions and then review the dispersive analysis used in this paper for deriving the non-forward positivity bounds. \\ \\ 
We work with the mostly minus signature of the Minkowski metric
\begin{equation}
    \eta_{\mu\nu}=\text{diag}(+1,-1,-1,-1)    
\end{equation}
A general momentum is given in the usual spherical polar coordinates as 
\begin{equation}
    p^\mu = (E, p\sin\theta\cos\phi, p\sin\theta\sin\phi, p\cos\theta)
\end{equation}
where $p\equiv |\Vec{p}|$. Here, $E^2=p^2+m^2$ and $m$ is the mass of external (spin-3/2) particles.  We will compute $2\to 2$ scattering amplitudes 
for scattering of identical, massive,  Majorana spin-3/2 particles
\begin{equation}
    \psi_{\mu}(p_1, \lambda_1)~ \psi_{\mu}(p_2,\lambda_2) \rightarrow \psi_{\mu}(p_3, \lambda_3)~ \psi_{\mu}(p_4,\lambda_4)    
\end{equation}
 in the centre of mass (COM) frame (i.e. $\Vec{p}_1+\Vec{p}_2=\Vec{p}_3+\Vec{p}_4=0$). Here, $p_1,~ p_2$ are the incoming and $p_3,~p_4$ are the outgoing 4-momenta, and $\lambda_i$ are the corresponding helicities of the external states. \\ \\The angles for the external states are as follows:
\begin{align}
    \theta_1=0, ~\phi_1=\varphi~;~~\theta_2=\pi-0, ~\phi_2=\varphi+\pi \\[4pt]
\theta_3=\theta, ~\phi_3=\varphi~;~~\theta_4=\pi-\theta, ~\phi_4=\varphi+\pi
\end{align}
In terms of the Mandelstam variables 
\begin{equation}
    s\equiv (p_1+p_2)^2,~t\equiv(p_3-p_1)^2,~u\equiv (p_4-p_1)^2
\end{equation}
we denote the amplitude 
\begin{equation}
     \bra{p_3,\lambda_3;p_4,\lambda_4}\hat T\ket{p_1,\lambda_1;p_2,\lambda_2} =(2\pi)^4\delta^{(4)}(0) \mathcal{M}_{\lambda_1\lambda_2}^{\lambda_3\lambda_4}(s,t)
\end{equation}where ${\hat S}=1+i{\hat T}$. \\\\
We will follow the conventions of \cite{Christensen:2013aua} for the expressions of polarization vectors of spins-1/2, 1 and 3/2.
The Dirac spin-1/2 $u$,$v$ spinors in the Weyl basis of gamma matrices are
\begin{equation}
\begin{aligned}
    u_+ = \begin{pmatrix}
        \sqrt{E-p}\cos(\theta/2)\\  \sqrt{E-p}\sin(\theta/2)e^{i\phi}\\  \sqrt{E+p}\cos(\theta/2)\\\sqrt{E+p}\sin(\theta/2)e^{i\phi}
    \end{pmatrix}  ,\quad u_- = \begin{pmatrix}
        -\sqrt{E+p}\sin(\theta/2)e^{-i\phi}\\  \sqrt{E+p}\cos(\theta/2)\\ - \sqrt{E-p}\sin(\theta/2)e^{-i\phi}\\\sqrt{E-p}\cos(\theta/2)
    \end{pmatrix} 
\end{aligned}
\end{equation}
The corresponding $v$-spinors can be obtained by the relation (true in Weyl basis) 
\begin{equation}v=-i\gamma^2u^*
\end{equation}
The spin-1 transverse and longitudinal polarizations are respectively:
\begin{align}
    \epsilon^\mu_\lambda(\Vec{p}) &= -\frac{\lambda}{\sqrt{2}}(0,~\cos\theta\cos\phi -i\lambda\sin\phi , \cos\theta\sin\phi +i\lambda\cos\phi , -\sin\theta) \label{spin1transversepols}\\
    \epsilon^\mu_0(\Vec{p}) &= \frac{1}{m}(p,E\sin\theta\cos\phi, E\sin\theta\sin\phi,E\cos\theta)
\end{align}
Spin-3/2 polarizations are given by the following linear combinations of products of spin-1 and spin-1/2 polarizations.
\begin{align}
    U^\mu_{+}(p) &= \sqrt{\frac{2}{3}}\, \epsilon^\mu_{0} u_{+} + e^{i\phi}\sqrt{\frac{1}{3}}\, \epsilon^\mu_{+} u_{-} \\[4pt]
    U^\mu_{-}(p) &= e^{i\phi}\sqrt{\frac{2}{3}}\, \epsilon^\mu_{0}u_{-} + \sqrt{\frac{1}{3}}\, \epsilon^\mu_{-} u_{+}\\[4pt]
    U^\mu_{(++)}(p) &= \epsilon^\mu_{+} u_{+} \\[4pt]
    U^\mu_{(--)}(p) &= e^{i\phi}\epsilon^\mu_{-}u_{-} 
\end{align}
 Here, $+$ and $-$ denote the longitudinal states of $\psi_\mu$ with helicities $\lambda = +1/2$ and $\lambda = -1/2$ respectively, and $(++)$ and $(--)$ denote the transverse $\lambda=+3/2$ and $\lambda=-3/2$ states respectively.\\ \\
In the Weyl Basis, $V^\mu(p) = -i\gamma^2 U^{\mu*}(p)$, so
\begin{align}
    V^\mu_{+}(p) &= \sqrt{\frac{2}{3}}\, \epsilon^{*\mu}_{0} v_{+} + e^{-i\phi}\sqrt{\frac{1}{3}}\, \epsilon^{*\mu}_{+} v_{-}\\[4pt]
    V^\mu_{-}(p) &= e^{-i\phi}\sqrt{\frac{2}{3}}\, \epsilon^{*\mu}_{0}v_{-} + \sqrt{\frac{1}{3}}\, \epsilon^{*\mu}_{-} v_{+}\\[4pt]
    V^\mu_{(++)}(p) &= \epsilon^{*\mu}_{+} v_{+} \\[4pt]
    V^\mu_{(--)}(p) &= e^{-i\phi}\epsilon^{*\mu}_{-}v_{-} 
\end{align}
With these conventions, the momentum space Equations of Motion \cite{Moroi:1995fs} are automatically satisfied on-shell (when $E^2=p^2+m^2$).
\begin{equation}\label{EOM}
    p^\mu U_\mu=0,~~\gamma^\mu U_\mu = 0,~~(\gamma^\mu p_\mu -m)U_\nu = 0
\end{equation}
We compute the contact contributions to the amplitudes using the Feynman Rules derived in Appendix \ref{Feynman Rules Derivation}. The graviton exchange results are taken from \cite{Antoniadis:2022jjy}.\\ \\
We will use the following result \cite{Bellazzini:2023nqj_MassiveGravity, Bellazzini:2025shd}, the derivation of which is reviewed in Appendix \ref{ArcAnalysis}.
Consider the Arc $\mathcal{A}$ defined by the contour integral over $\mathcal{C}$ as shown in Figure \ref{Contour}.
\begin{equation}\label{ArcDefinition}
    \mathcal{A}_{\lambda_1\lambda_2}^{\lambda_3\lambda_4}(t,n)=\oint_\mathcal{C} \frac{ds}{2\pi i}\frac{\mathcal{M}_{\lambda_1\lambda_2}^{\lambda_3\lambda_4}+\mathcal{M}_{\lambda_1\bar\lambda_4}^{\lambda_3\bar\lambda_2}}{\left(s-2m^2+\frac{t}{2}\right)^{3+n}}
 \end{equation}
 Then it follows from Unitarity and analyticity properties of the amplitudes, that the non-forward arcs are bounded by the forward ones:
\begin{equation}\label{ArcResult}
    \frac{|\mathcal{A}_{++}^{++}(t,0)|}{\mathcal{A}_{++}^{++}(0,0)}\leq 1 + \mathcal{O}\left(\frac{\sqrt{-t}m}{\Lambda^2}\right)~~~\&~~~\frac{|\mathcal{A}_{++}^{--}(t,0)|+|\mathcal{A}_{+-}^{-+}(t,0)|}{2 \mathcal{A}_{++}^{++}(0,0)}\leq 1 + \mathcal{O}\left(\frac{\sqrt{-t}m}{\Lambda^2}\right)
\end{equation}
The corrections on RHS come due to the non-trivial crossing symmetry in the amplitudes for massive particles. See for example, \cite{Hebbar:2020ukp, Bellazzini:2023nqj_MassiveGravity, Bellazzini:2025shd}. These can be ignored by working in a regime where $|t|$ and $m$ are small enough compared to $\Lambda$; see Appendix \ref{Order_of_Corrections}.
  
\section{Majorana Spin-3/2 EFTs}\label{Only Contact}
To illustrate the general idea and motivate the coupling to gravity, we first consider only mass dimension-6 contact operators, and derive non-forward positivity bounds by applying the Arc analysis Eq. (\ref{ArcResult}) to $2\to 2$ longitudinal scattering of massive, Majorana Spin-3/2 particles, considering only contact interactions.  We start with only one operator, giving the general idea in section \ref{Fermi_EFT}, and then generalize to three contact operators in section \ref{dim6_EFT}. \\\\ Some of the results presented in this section were previously derived in \cite{Bellazzini:2025shd}. We nevertheless reproduce them here in our notation to keep the discussion self-contained and to facilitate the connection with the analysis of section~\ref{Neighbourhood of Supergravity}.
\subsection{Spin-3/2 Fermi theory}\label{Fermi_EFT}
We first consider the spin-3/2 analogue of the Fermi interaction, and derive positivity constraints on its coupling.
\begin{equation}
    \mathcal{L}_{\mathrm{int}}=\frac{k_S}{\Lambda^2}(\bar\psi_\mu\psi^\mu)^2
\end{equation}
  The $s$-$u$ symmetric (up to crossing corrections, as explained in Appendix \ref{ArcAnalysis}) amplitudes due to this interaction are given by
 \begin{align}
    \mathcal{M}_{++}^{++}+\mathcal{M}_{+-}^{+-} &= -\frac{8k_S}{\Lambda^2}\frac{t(3s^2+3st+t^2)}{9m^4}+\mathcal{O}(E^4/m^2\Lambda^2)  \\[4pt]
    \mathcal{M}_{++}^{--} &= -\frac{8k_S}{\Lambda^2}\frac{st(s+t)}{3m^4} +\mathcal{O}(E^4/m^2\Lambda^2)\\[4pt]
    \mathcal{M}_{+-}^{-+} &= -\frac{8k_S}{\Lambda^2}\frac{t^3}{9m^4}+\mathcal{O}(E^4/m^2\Lambda^2)
\end{align}
Only the $E^6$ contributions are mentioned for simplicity. 
The $E^4$ contributions are given in Appendix \ref{Full Amplitudes}. 
The Arcs are  computed according to Eq. (\ref{ArcDefinition}), including the $E^4$ pieces and hence for arbitrary $m$. 
\begin{align}
     \mathcal{A}_{++}^{++} &= -\frac{8}{9m^4\Lambda^2}\left(3k_St +8k_Sm^2\right) \\[4pt]
     \mathcal{A}_{++}^{--} &= -\frac{16}{9m^4\Lambda^2}\left(3k_St +8k_Sm^2\right)=2\mathcal{A}_{++}^{++} \\[4pt]
     \mathcal{A}_{+-}^{-+} &=0
 \end{align}
Then Arc analysis Eq. (\ref{ArcResult}) gives two constraints. First, the positivity of forward elastic arc requires
\begin{equation}
    k_S<0
\end{equation}
Interestingly, this rules out the Yukawa Scalar UV completion since it gives a positive coupling on integrating out the heavy scalar\footnote{Apart from deriving Feynman rules from scratch (see Appendix \ref{Feynman Rules Derivation}), to fix the sign, we have done several checks to make sure there is no overall sign mistake. One of them is computing massive scalar exchange amplitudes, with $M_{\Phi}\to \infty$ giving the contact (Fermi theory) amplitudes and $M_{\Phi}\to 0$ giving the $-E^4$ growing amplitudes which match with the literature \cite{Bellazzini:2025shd, Gherghetta:2024tob, Gherghetta:2025tlx}. A check on the Feynman rules themselves is that in the spin-1/2 case ($U_\mu \to u$), they correctly reduce to those in \cite{Hebbar:2020ukp}. }: 
\begin{equation}
    -\frac{1}{2}\Phi(\Box +M_{\Phi}^2)\Phi+\frac{g}{2}\bar\psi_\mu\psi^\mu\Phi\xrightarrow{M_{\Phi}\gg E}+\frac{g^2}{8M_{\Phi}^2}(\bar\psi_\mu\psi^\mu)^2
\end{equation}
If $k_S<0$ is satisfied, we have further constraints due to the non-forward arcs:
\begin{equation}
    |\mathcal{A}_{++}^{++}(t,0)|\leq \mathcal{A}_{++}^{++}(0,0)~~~\&~~~|\mathcal{A}_{++}^{--}(t,0)|+|\mathcal{A}_{+-}^{-+}(t,0)|\leq 2 \mathcal{A}_{++}^{++}(0,0)
\end{equation}
Both give
   \begin{equation}
      \left|1+\frac{3t}{8m^2}\right|\leq 1 
  \end{equation}
 In the regime where both crossing and higher derivative corrections are small, this implies a constraint on the range of $t$:
 \begin{equation}
     -16m^2/3<t<0
 \end{equation}
 which is inconsistent with the larger allowed kinematical range $t$ can have (under our assumptions): \begin{equation}
     -\Lambda^2\ll t<0
 \end{equation}
 unless $16m^2/3\geq |t|_{\rm max}$ or
 \begin{equation}\label{m_bound}
     \frac{16m^2}{3}\geq \frac{\Lambda^2}{N^2} \implies \frac{m^2}{\Lambda^2}\geq \frac{3}{16N^2}
 \end{equation}
 where $|t|_{\rm max}\equiv \Lambda^2/N^2$ is the maximum value of $|t|$ such that crossing corrections remain negligible.\\ \\ We must choose $N$ appropriately, to minimize crossing-corrections in Eq. (\ref{ArcResult}).
 Taking $N$ very large minimizes  the crossing corrections but just corresponds to  the forward ($t\approx 0$) arc positivity and is not very useful. So we choose $N = 10$, which turns out to be a good compromise between strength of the bound and its accuracy. There are also potential corrections from higher derivative terms, which are discussed in Appendix \ref{Order_of_Corrections}, which suggests that for $m<(0.3)\Lambda$, the bounds can be trusted to within $10\%$ accuracy.\\ \\
In this simple example, the allowed region in the EFT parameter (sub)space is
 \begin{equation}\label{FermiBound}
     k_S<0 \text{~~and~~} m \gtrsim (0.04)\Lambda
 \end{equation}
 This example makes clear why the spin-3/2 mass has a lower bound in terms of the EFT cut-off. The same conclusion holds true when one adds more contact operators, as we will show in section \ref{dim6_EFT}. \\ \\
 We can also examine if adding a light\footnote{By ``light", we mean that the mass of scalar $M_S$ is smaller than the radius of the contour in Figure \ref{Contour}, set by the centre of mass energy $s$. Operationally, this is relevant when one Taylor expands denominators of the form 
 \begin{equation}
     \frac{1}{s-M_S^2} \approx \frac{1}{s}\left(1+\frac{M_S^2}{s}+...\right)
 \end{equation} \label{light}} scalar through relevant interactions in the IR makes it possible to allow for arbitrarily small mass in the EFT parameter space. Consider
 \begin{equation}
     \mathcal{L} \supset \frac{k_S}{\Lambda^2}(\bar\psi_\mu\psi^\mu)^2 + \frac{1}{2}g_S \bar\psi_\mu\psi^\mu\phi
 \end{equation}
 The light scalar contributes to the amplitudes at $\mathcal{O}(E^4)$:
 \begin{align}
    \mathcal{M}_{++}^{++}+\mathcal{M}_{+-}^{+-} &= -g_S^2\frac{2s^2+2st+t^2}{9m^4} \\[4pt]
    \mathcal{M}_{++}^{--} &= -2g_S^2\frac{s^2+st+t^2}{9m^4} \\[4pt]
    \mathcal{M}_{+-}^{-+} &= g_S^2\frac{t^2}{9m^4}
\end{align}
After including these contributions, the corresponding (total) arcs are
\begin{align}
     \mathcal{A}_{++}^{++} &=-\frac{2g_S^2}{9m^4}-\frac{8}{9m^4\Lambda^2}\left(3k_St +8k_Sm^2\right) \\[4pt]
     \mathcal{A}_{++}^{--} &= -\frac{4g_S^2}{9m^4}-\frac{16}{9m^4\Lambda^2}\left(3k_St +8k_Sm^2\right) \\[4pt]
     \mathcal{A}_{+-}^{-+} &= 0
 \end{align}
 We see that the light scalar gives negative contribution to the forward-elastic arc, and hence results in a further constrained space of allowed parameters. The bounds Eqs. (\ref{m_bound}) and (\ref{FermiBound}) get modified to
 \begin{equation}
     k_S<-\frac{g_S^2\Lambda^2}{32m^2},~~\frac{m^2}{\Lambda^2}>\frac{3}{16N^2}+\frac{g_S^2}{32|k_S|}
 \end{equation}
 Thus, addition of light scalar results in stronger constraints on the parameters. In particular, it \emph{increases} the lower bound on the mass $m$. 

 \subsection{General dim-6 EFT}\label{dim6_EFT}
\begin{figure}[h!]
    \centering
    \includegraphics[width=0.4\linewidth]{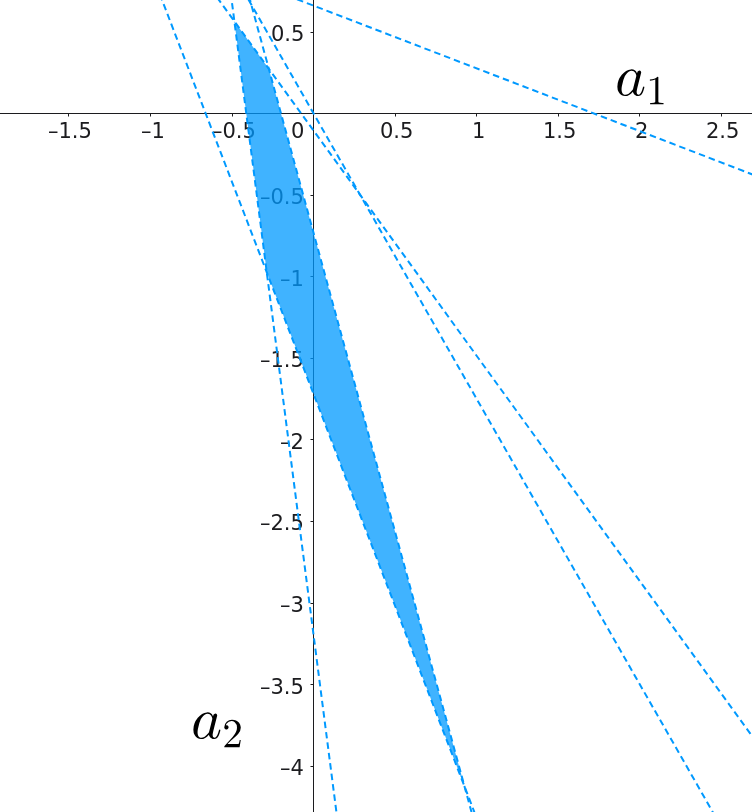}
    \caption{For fixed $A=4.5 ~(m\sim\Lambda/21.2),~~a_+=-0.2$, we get a bounded region in $a_1$-$a_2$ coupling space. But in the 3D $a_1$-$a_2$-$a_+$ space, this is still unbounded in the $a_+$ direction.}
    \label{fig:only_contact_bounded}
\end{figure}

 We now generalize the analysis of the previous subsection to the case with three dim-6 operators:
 \begin{equation}\label{OnlyContactL}
     \mathcal{L}_{\mathrm{int}}=\frac{(k_S+k_P)}{2\Lambda^2}\left((\bar\psi_\mu\psi^\mu)^2+(\bar\psi_\mu\gamma^5\psi^\mu)^2\right)+\frac{ a_1}{\Lambda^2}(\bar\psi^\mu \gamma^\rho \psi^\nu)(\bar\psi_\mu \gamma_\rho \psi_\nu)+\frac{ a_2}{\Lambda^2}(\bar\psi^\rho \gamma^\mu \psi^\nu)(\bar\psi_\rho \gamma_\nu \psi_\mu)
 \end{equation}
 This is equivalent to the basis considered in \cite{Bellazzini:2025shd}:
 \begin{equation}
     \mathcal{L}_{\mathrm{int}}=\frac{k_S}{\Lambda^2}(\bar\psi_\mu \psi^\mu)^2+\frac{k_V}{\Lambda^2}(\bar\psi_\mu \gamma^\alpha \gamma^5\psi^\mu)^2 + \frac{k_P}{\Lambda^2}(\bar\psi_\mu \gamma^5 \psi^\mu)^2
 \end{equation}
 which can  be rewritten as 
  \begin{align}
     \mathcal{L}_{\mathrm{int}}&=\frac{(k_S+k_P)}{2\Lambda^2}\left((\bar\psi_\mu\psi^\mu)^2+(\bar\psi_\mu\gamma^5\psi^\mu)^2\right)+\frac{(k_S-k_P)}{2\Lambda^2}\left((\bar\psi_\mu\psi^\mu)^2-(\bar\psi_\mu\gamma^5\psi^\mu)^2\right) \nonumber\\[4pt]
     &+ \frac{k_V}{\Lambda^2}(\bar\psi_\mu \gamma^\alpha \gamma^5\psi^\mu)^2 \label{basis2}
 \end{align}
 We define for simplicity
 \begin{equation}
     a_+ = \frac{k_S+k_P}{2}
 \end{equation}
 The rest of the coefficients in the two bases Eq. (\ref{OnlyContactL}) and Eq. (\ref{basis2}) are related as
 \begin{equation}
     a_1=(k_S-k_P)+2k_V,~~~a_2=-(k_S-k_P)-4k_V
 \end{equation}
 We will work with the basis Eq. (\ref{OnlyContactL}) so that it is easier to generalize to the SUGRA case Eq. (\ref{SUGRAEFT}). The relevant longitudinal amplitudes are 
 \begin{align}
    \mathcal{M}_{++}^{++} &= \frac{8}{9m^4\Lambda^2}\left(a_1s^3+2(a_1+a_2)stu\right)+\mathcal{O}(E^4/m^2\Lambda^2)\\[4pt]
    \mathcal{M}_{+-}^{+-} &= \frac{8}{9m^4\Lambda^2}\left(a_1u^3+2(a_1+a_2)stu\right)+\mathcal{O}(E^4/m^2\Lambda^2)\\[4pt]
    \mathcal{M}_{+-}^{-+} &= \frac{-8}{9m^4\Lambda^2}\left(a_1t^3+2(a_1+a_2)stu\right)+\mathcal{O}(E^4/m^2\Lambda^2)\\[4pt]
    \mathcal{M}_{++}^{--} &=\frac{16}{3m^4\Lambda^2}a_+ stu -\frac{32}{9m^2\Lambda^2}\left((3a_1+a_2)s^2-(9a_1+5a_2)tu\right)+\mathcal{O}(E^2/\Lambda^2)
\end{align}\\
The full amplitudes including the $E^4$ contributions are given in Appendix \ref{Full Amplitudes}.
The corresponding Arcs are
 \begin{align}
     \mathcal{A}_{++}^{++} &= -\frac{8}{9m^4\Lambda^2}\left((7a_1+4a_2)t+8(a_1+a_2+2a_+)m^2\right) \\[4pt]
     \mathcal{A}_{++}^{--} &= -\frac{8}{9m^4\Lambda^2}\left(12a_+t+8(3a_1+a_2)m^2\right)  \\[4pt]
     \mathcal{A}_{+-}^{-+} &= -\frac{8}{9m^4\Lambda^2}\left(-4(a_1+a_2)t+0m^2\right) 
 \end{align}
Arc analysis gives the following:
\begin{eqnarray}
    2(5a_1+3a_2+4a_+)m^2&<(a_1+a_2-3a_+)t<&2(a_1-a_2-4a_+)m^2 \\[4pt]
    -2(a_1-a_2-4a_+)m^2&<(a_1+a_2+3a_+)t<&-2(5a_1+3a_2+4a_+)m^2
\end{eqnarray}
Setting $t=-\Lambda^2/N^2$ and defining
\begin{equation}
    A\equiv \frac{\Lambda^2}{N^2m^2}
\end{equation}
we have the following constraints:
\begin{align}
    (10+A)a_1+(6+A)a_2&<(3A-8)a_+ \label{ineq1}\\[4pt] 
    (2+A)a_1+(-2+A)a_2 &> (3A+8)a_+ \\[4pt]
    (-2+A)a_1+(2+A)a_2 &< -(3A+8)a_+ \\[4pt]
    (-10+A)a_1+(-6+A)a_2 &> -(3A-8)a_+ \\[4pt]
    (16-7A)a_1+4(4-A)a_2&<-32a_+ \\[4pt]
    7a_1+4a_2&<0 \label{ineq6}
\end{align}
This leads to an upper bound on $A$ and hence a lower bound on $m$:
\begin{equation}\label{masscutoff}
    A\lesssim 5.3\implies \frac{N^2m^2}{\Lambda^2}\gtrsim\frac{1}{5.3} \implies m>\frac{\Lambda}{\sqrt{5.3}N}\sim (0.04)\Lambda
\end{equation}
For $A\geq 5.3$ the region simply vanishes. There is no smooth $m\to 0$ limit in this case i.e. the spin-3/2 mass is not allowed to be arbitrarily small. If $A<5.3$ is satisfied, in general, we get an unbounded region. A plot with $A=3$ $(m\sim \Lambda/17.3)$ and $a_+=0.02$ is shown below in Figure \ref{fig:only_contact_unbounded}. The 6 lines represent the 6 inequalities Eqs. (\ref{ineq1})-(\ref{ineq6}) obtained from Arc analysis and the shaded region is the space of allowed values of $a_1$ and $a_2$ for fixed $A=3,~a_+ = 0.02$. 
\begin{figure}[h!]
    \centering
    \includegraphics[width=0.4\linewidth]{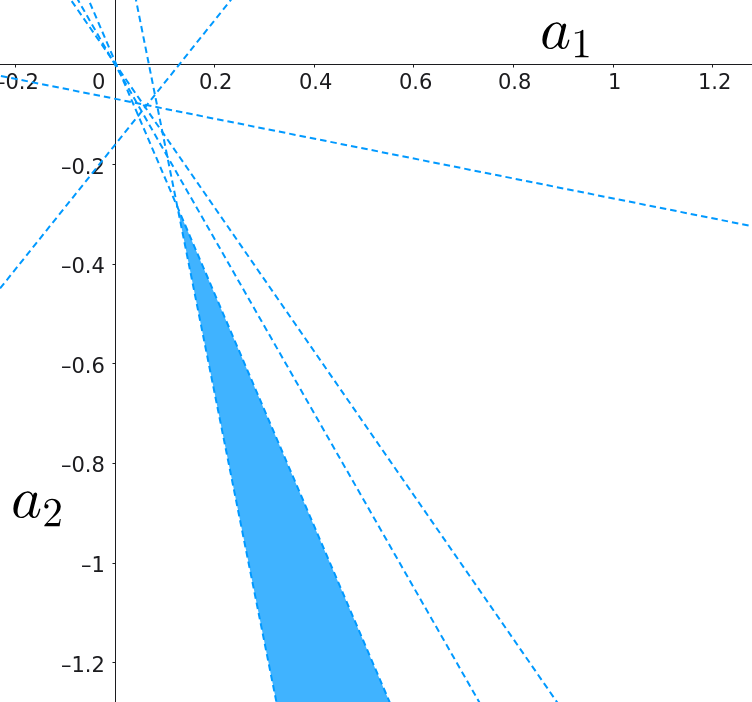}
    \caption{The space of allowed $a_1$ and $a_2$ for fixed $A=3~(m\sim \Lambda/17.3),~~a_+=0.02$.}
    \label{fig:only_contact_unbounded}
\end{figure} 
We get a region bounded in the $a_1$-$a_2$ subspace when $a_+<0$ and $4<A<5.3$ $(\Lambda/23 <m<\Lambda/20)$ (but still unbounded in the $a_+$ direction). See for example Figure \ref{fig:only_contact_bounded}.
\\
To conclude, when we consider only dim-6 contact couplings for the Majorana spin-3/2 EFT, the space of allowed contact couplings exists and is unbounded when the mass is above a certain finite cut-off Eq. (\ref{masscutoff}). Below this cutoff, the allowed region identically vanishes. \\ \\
The fact that the allowed region is unbounded is not surprising, since positivity only bounds ratios of Wilson coefficients, as opposed to partial wave unitarity which bounds absolute values. We showed the plots on absolute values here to compare them later with bounded regions found in section \ref{Neighbourhood of Supergravity} (where the graviton coupling is assumed to be fixed to minimal coupling) and from partial wave unitarity in section \ref{Partial Wave Unitarity}.

 \section{Neighbourhood of Supergravity}\label{Neighbourhood of Supergravity}
 We saw in the previous section that having only contact operators in the Majorana spin-3/2 EFT results in a lower bound on its mass in terms of the EFT cut-off, as also seen in \cite{Bellazzini:2025shd}. It was argued in \cite{Bellazzini:2025shd} that from positivity arguments, allowing arbitrarily small mass is possible only by adding a graviton into the theory. Furthermore, after adding the (minimally coupled) graviton, the dispersive analysis at leading order in $m\ll M_{\rm Pl}$ expansion  fixes the contact $\psi_\mu^4$ couplings to Supergravity values. In this section, we use the non-forward dispersive framework of \cite{Bellazzini:2025shd} keeping $m$ arbitrary, but smaller than $M_{\rm Pl}$, to see if a larger space of couplings is allowed in the neighbourhood of Supergravity point, once the (minimally coupled) graviton is added. We consider the Supergravity motivated Lagrangian\footnote{ We do not consider here contributions from gravitational multipoles, which are indeed shown to vanish at leading order by positivity arguments in \cite{Bellazzini:2025shd}. }
 \begin{equation}\label{SUGRAEFT}
     \mathcal{L}_{\mathrm{int}}=\frac{1}{M_{\rm Pl}}h_{\mu\nu}T^{\mu\nu}+\frac{a_+}{M_{\rm Pl}^2}\left((\bar\psi_\mu\psi^\mu)^2+(\bar\psi_\mu\gamma^5\psi^\mu)^2\right)+\frac{ a_1}{M_{\rm Pl}^2}(\bar\psi^\mu \gamma^\rho \psi^\nu)(\bar\psi_\mu \gamma_\rho \psi_\nu)+\frac{ a_2}{M_{\rm Pl}^2}(\bar\psi^\rho \gamma^\mu \psi^\nu)(\bar\psi_\rho \gamma_\nu \psi_\mu)
 \end{equation}
 and derive non-forward positivity bounds via Eq. (\ref{ArcResult}) on $a_1,~a_2~\&~a_+$, keeping $m$ arbitrary. Here, $T_{\mu\nu}$ is the energy-momentum tensor corresponding to massive Rarita-Schwinger field minimally coupled to gravity.
 In $\mathcal{N}=1$ Supergravity, the couplings are fixed to be \cite{Antoniadis:2022jjy}
 \begin{equation}\label{SUGRApoint}
     a_1 = -1/32,~~a_2 = -1/16,~~a_+=0
 \end{equation}
 The Supergravity Lagrangian contains one more contact $\psi_\mu^4$ operator: 
 \begin{equation}
     \frac{ a_3}{M_{\rm Pl}^2}(\bar\psi^\rho \gamma^\mu \psi_\mu)(\bar\psi_\rho \gamma^\nu \psi_\nu);~~~a_3=1/8
 \end{equation}
 which does not contribute at tree-level due to the equation of motion
 \begin{equation}
     \gamma^\mu \psi_\mu = 0
 \end{equation}
 and hence will not be considered in our analysis. So whenever we say ``Supergravity point", we mean Eq. (\ref{SUGRApoint}), with $a_3$ arbitrary.
 \subsection{Obstruction from $s^2/t$ pole}\label{t-pole}
 We first address the inevitable obstruction from the $s^2/t$ pole, coming from graviton exchange diagrams, in the following analysis.
 Due to the graviton exchange, the elastic amplitudes (and hence the corresponding arc) are not well defined in the forward limit. The relevant amplitude (needed to define $\mathcal{A}_{++}^{++}$) is schematically of the form
\begin{equation}
    \mathcal{M}_{++}^{++}+\mathcal{M}_{+-}^{+-} = \frac{E^6}{m^4M_{Pl}^2} + \frac{E^4}{m^2M_{Pl}^2} + ... + \frac{1}{M_{Pl}^2}\frac{s^2}{t}
\end{equation}
and hence the arc will be of the form
\begin{equation}
    \mathcal{A}_{++}^{++}(t) = \frac{t+m^2}{m^4M_{Pl}^2} + \frac{1}{t\,M_{Pl}^2}
\end{equation}
Even if the limit $m\to 0$ is taken first, so that the $t$-pole in the last term can be killed,
we would still have a problem because this would translate to $m^2\ll |t|$; but once this limit is taken, the  arc is of the form
\begin{equation}
    \mathcal{A}_{++}^{++}(t) = \frac{t+m^2}{m^4 M_{\rm Pl}^2}
\end{equation}
and ``taking $t\to 0$" translates to $|t|\ll m^2$ which contradicts the former limit $m^2\ll |t|$. To see this clearly, it is best to make everything dimensionless by considering the arc $m^2M_{\rm Pl}^2 \mathcal{A}_{++}^{++}$:
\begin{equation}
    m^2M_{\rm Pl}^2 \mathcal{A}_{++}^{++}(t) = \frac{t}{m^2}+1+\frac{m^2}{t}
\end{equation}
Then we see that there is just one independent variable $t/m^2$ and only one limit can be taken.
To neglect the last term, one needs $m^2\ll |t|$ and then to define the forward limit, one needs $|t|\ll m^2$.
Working in the decoupling limit
would simply remove the graviton from the theory, and the analysis would reduce to that of the previous section. There has been progress to solve this issue in space-time dimensions $d\geq 5$ \cite{Caron-Huot:2021rmr,Beadle:2024hqg, Beadle:2025cdx}, by working with smeared dispersion relations, but extending their analysis to $d=4$ is not yet clear.
See \cite{Bellazzini:2025bay} for recent work on bounds in $d=4$. Although formally one cannot ignore the $t$-channel pole, it might be possible to make its contribution sub-leading in the regime $m\ll M_{\rm Pl}$ by making use of these smeared dispersion relations. See the discussion around Eq. (\ref{IR_log}), where an argument is made in the context of partial wave analysis, which may be applied to smeared dispersive analyses, and neglecting the $t$-pole can be made more rigorous in the regime $m\ll M_{\rm Pl}$, but such an argument would not hold when $m\sim M_{\rm Pl}$. We leave this for future consideration.\\ \\
For now, the most convenient way to proceed is to subtract the $s^2/t$ pole, which we treat as an assumption, and perform the analysis. This is the prescription we will be using throughout the rest of the paper to define the forward limits. This procedure is of course questionable, but is often used in the literature. See for instance \cite{Cheung:2014ega}. Interestingly, they show that the bounds that follow after this ad-hoc $t$-pole subtraction, also follow from independent causality and unitarity arguments. 
\subsection{Analysis}
\begin{figure}[h!]
    \centering
    \includegraphics[width=0.5\linewidth]{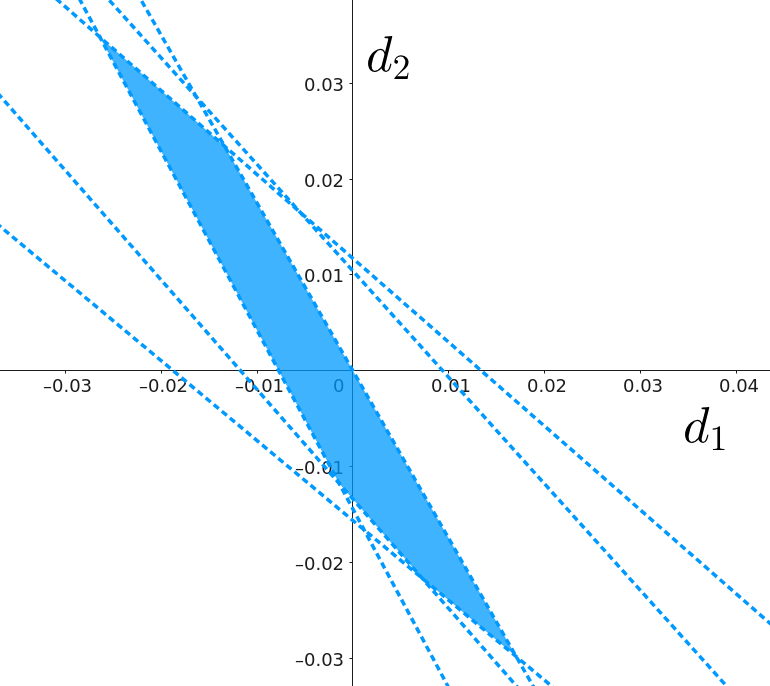}
    \caption{The shaded region is the space of allowed deviations from ``SUGRA point" $(0,0)$ in the $d_2$-$d_1$ subspace with  $a_+=0$ and $A=30~(m\sim M_{\rm Pl}/54.8)$.}
    \label{fig:d1-d2-0}
\end{figure}

 After removing the $s^2/t$ terms from the amplitudes, the analysis can be carried out similar to the previous section.
 The relevant longitudinal scattering amplitudes are: 
  \begin{align}
    \mathcal{M}_{++}^{++} &= \frac{s^3+6stu}{36m^4M_{\rm Pl}^2}+\frac{8}{9m^4\Lambda^2}\left(a_1s^3+2(a_1+a_2)stu\right)+\mathcal{O}(E^4/m^2M_{\rm Pl}^2) \label{M++++}\\[4pt]
    \mathcal{M}_{+-}^{+-} &=  \frac{u^3+6stu}{36m^4M_{\rm Pl}^2}+\frac{8}{9m^4\Lambda^2}\left(a_1u^3+2(a_1+a_2)stu\right)+\mathcal{O}(E^4/m^2M_{\rm Pl}^2)\\[4pt]
    \mathcal{M}_{+-}^{-+} &=-\frac{t^3+6stu}{36m^4M_{\rm Pl}^2} -\frac{8}{9m^4\Lambda^2}\left(a_1t^3+2(a_1+a_2)stu\right)+\mathcal{O}(E^4/m^2M_{\rm Pl}^2)\\[4pt]
    \mathcal{M}_{++}^{--} &=\frac{16}{3m^4\Lambda^2}a_+ stu-\frac{1}{9m^2\Lambda^2s}\left(5s^2+19st+19t^2\right) \label{M++--} \\[4pt]
    & -\frac{32}{9m^2\Lambda^2}\left((3a_1+a_2)s^2-(9a_1+5a_2)tu\right)+\mathcal{O}(E^2/M_{\rm Pl}^2)\nonumber
\end{align}
The graviton exchange contributions are taken from \cite{Antoniadis:2022jjy}.
The $E^4$ contributions in the amplitudes are omitted here for simplicity; they are given in Appendix \ref{Full Amplitudes}. The corresponding exact arcs (including the $E^4$ contributions) are
\begin{align}
     \mathcal{A}_{++}^{++} &= -\frac{1}{36m^4M_{\rm Pl}^2}\left((15+224a_1+128a_2)t+256(2a_+ + a_1 +a_2)m^2\right) \\[4pt]
     \mathcal{A}_{++}^{--} &= -\frac{2}{9m^4M_{\rm Pl}^2}\left(48a_+t+(5+96a_1+32a_2)m^2\right)\\[4pt]
     \mathcal{A}_{+-}^{-+} &= \frac{t}{9m^4}\left(3+32a_1+32a_2\right)
 \end{align}
We rewrite the arcs in terms of deviations from SUGRA values:
\begin{equation}
    a_1=-\frac{1}{32}+d_1,~~a_2=-\frac{1}{16}+d_2,~~a_+=d_+
\end{equation}
The arcs in terms of $d_1,~ d_2, ~a_+$ are
\begin{align}
    \mathcal{A}_{++}^{++} &= -\frac{2}{9m^4M_{\rm Pl}^2}\left(4(7d_1+4d_2)t+(32d_1+32d_2+64a_+-3)m^2\right) \\[4pt]
    \mathcal{A}_{++}^{--} &= -\frac{32}{9m^4M_{\rm Pl}^2}\left(3a_+t+(6d_1+2d_2)m^2\right)\\[4pt] 
    \mathcal{A}_{+-}^{-+} &=\frac{32}{9m^4M_{\rm Pl}^2}\left(d_1+d_2\right)t
\end{align}
 Arc analysis Eq. (\ref{ArcResult}) gives the following constraints:
\begin{align}
    (10+A)d_1+(6+A)d_2-3/8&<(3A-8)a_+ \label{bound1} \\[4pt] 
    (2+A)d_1+(-2+A)d_2+3/8 &> (3A+8)a_+ \\[4pt]
    (-2+A)d_1+(2+A)d_2 -3/8 &< -(3A+8)a_+ \\[4pt]
    (-10+A)d_1+(-6+A)d_2 + 3/8&> -(3A-8)a_+ \\[4pt]
    (16-7A)d_1+4(4-A)d_2-3/2&<-32a_+ \\[4pt]
    7d_1+4d_2&<0 \label{bound6}
\end{align}
where we define $A\equiv M_{\rm Pl}^2/N^2m^2$.
A plot of the allowed region in $d_2$-$d_1$, $a_+=0$ subspace, for $A=30$ is shown in Figure \ref{fig:d1-d2-0} as an example. The 6 lines correspond to the 6 inequalities Eqs. (\ref{bound1})-(\ref{bound6}) and the shaded region is their intersection. We see that the SUGRA point ($d_1=d_2=a_+=0$) lives on one of the boundaries ($7d_1+4d_2<0$) of this region.
Interestingly, the couplings are bounded from all sides, forming a closed polytope in the $d_1$-$d_2$ space. This can be seen analytically if one works in the small $m$ approximation. 
\subsection{Small mass approximation}
\begin{figure}[h!]
    \centering
    \includegraphics[width=0.5\linewidth]{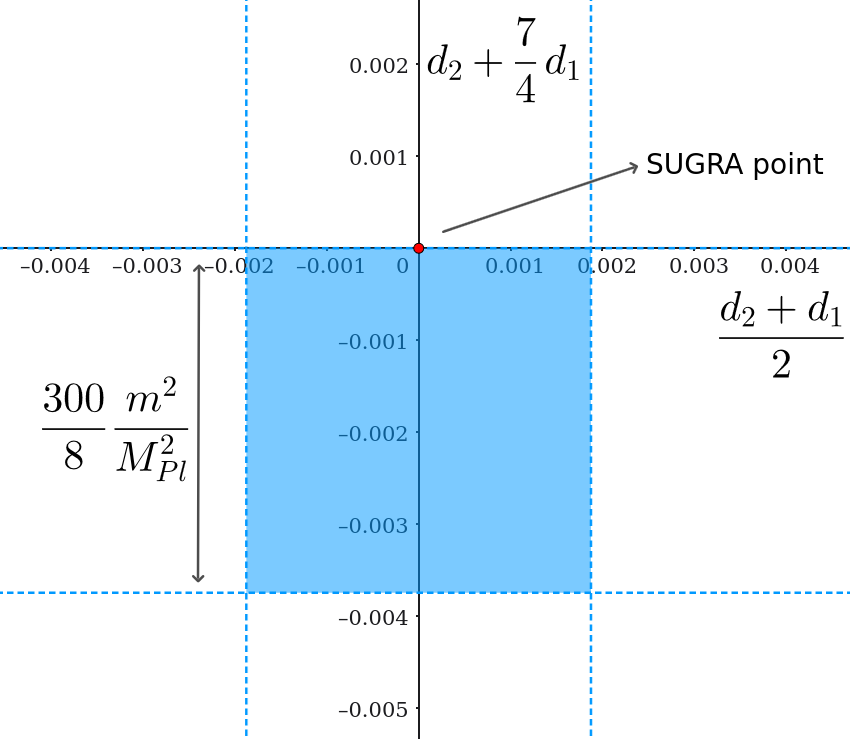}
    \caption{For $a_+=0$, $A=100~(m= M_{\rm Pl}/100)$ the allowed region is a square with side $3\big/8A$, with the Supergravity point $(0,0)$ at the top edge. As $m$ approaches $0$, so does $A^{-1}$ and hence the square shrinks to the origin.}
    \label{Square}
\end{figure}

The equations simplify significantly in the small mass or large $A$ approximation i.e. $d_iA\gg d_i$ (or $A\gg 10$ due to Eq. (\ref{bound1})) and we can get an analytical expression of the allowed region and its area/volume. For $N=10$, $A\gg 10$  translates to $m\ll M_{\rm Pl}/31.6$, which is a very reasonable assumption. Even for $m\sim$ TeV, $m/M_{\rm Pl}\sim 10^{-16}$ which fits well within this window. \\ \\
In the $a_+=0$ subspace, the bounded region is approximated by 
\begin{align}
    -\frac{3}{8A}&<d_2+\frac{7}{4}d_1<0 \label{square_1}\\[4pt]
    -\frac{3}{8A}&<d_2+d_1~~<\frac{3}{8A} \label{square_2}
\end{align}
This is just a square with side $3/8A$ in $u$-$v$ space where $u\equiv \frac{1}{2}(d_2+d_1)$ and $v\equiv (d_2+\frac{7}{4}d_1)$, with the SUGRA point on the top edge, as shown in Figure \ref{Square}. Its area is just $(3/8A)^2$ in $u$-$v$ space. In $d_1$-$d_2$ space, by multiplying with the Jacobian factor ($8/3$), we get
\begin{equation}\label{positivity_area}
    \mathrm{Area}  = \frac{3}{8}(A^{-1})^2 = \frac{3N^4}{8}\frac{m^4}{M_{\rm Pl}^4}
\end{equation}
So we see explicitly that the region shrinks to the origin (SUGRA point) when $m\to 0$.
\subsubsection*{Volume of the allowed region}
Generalizing to arbitrary $a_+$, in the large $A$ (or small $m$) approximation, we see that the region is bounded from all sides:
\begin{equation}
    -\frac{3}{8A}<(d_1+d_2)\pm 3a_+<\frac{3}{8A}~~~\&~~~\frac{64a_+-3}{2A}<7d_1+4d_2<0
\end{equation}
The first two inequalities have a non-zero intersection only when $-1/8A<a_+<1/8A$. \\ \\ More explicitly, the intersection of the above inequalities gives
\begin{align}
    -\frac{1}{8A}&<a_+<\frac{1}{8A} \\[4pt]
    -\left(\frac{3}{8A}-3|a_+|\right) &< d_2+d_1 <\left(\frac{3}{8A}-3|a_+|\right) \\[4pt]
    \frac{64a_+-3}{8A} &< d_2+\frac{7}{4}d_1 <0
\end{align}
%
Interestingly, these linear inequalities form a polytope in the $d_1$-$d_2$-$a_+$ space due to being bounded from all sides. Such geometries in the context of EFTs had gained interest \cite{Arkani-Hamed:2020blm} due to their potential connection to the ``Amplituhedron" \cite{Arkani-Hamed:2013jha}.   In $d_1$-$d_2$-$a_+$ space, the volume of this polytope is
\begin{align}
    \mathrm{Vol} &= \frac{3}{64}\frac{1}{A^3} + \mathcal{O}(A^{-4}) \\[4pt]
    &= \frac{3N^6}{64}\frac{m^6}{M_{\rm Pl}^6}+\mathcal{O}(m^8/M_{\rm Pl}^8)\label{Volume}
\end{align}
Due to the last inequality $d_2+\frac{7}{4}d_1<0$, we see that for large $A$, the point $d_1=d_2=0$ lives on the boundary of this region even for $a_+\neq 0$.  This expression for volume can also be written in terms of $t$ (through the definition  $t=-M_{\rm Pl}^2/N^2$) as
\begin{equation}\label{Volume_t}
    \mathrm{Vol} = \frac{3}{64}\frac{m^6}{|t|^3}
\end{equation}
Note that the space of absolute values of Wilson coefficients is bounded because the gravitational  minimal coupling is fixed. If it is restored to an arbitrary coupling $(g/M_{\rm Pl})h_{\mu\nu}T^{\mu\nu}$,  it is easy to see from Eqs. (\ref{square_1}) and (\ref{square_2}) that 1) $g>0$ and 2) the space of ratios $d_i/g$ is the one that is bounded. Arbitrarily large $g$ is allowed from positivity, but can be constrained by partial wave unitarity. 
\subsection{Far from SUGRA regime}
\begin{figure}[h!]
    \centering
    \includegraphics[width=0.4\linewidth]{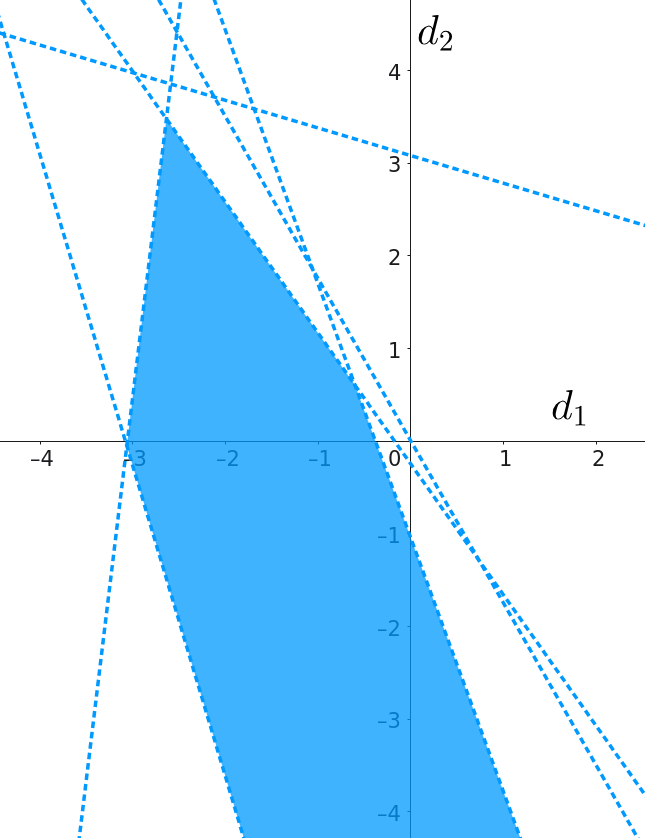}
    \caption{The allowed region is unbounded in the (negative) $d_2$ direction for $A=3.7~(m\sim M_{\rm Pl}/19.2),~a_+=-1$.}
    \label{fig:Unbounded}
\end{figure}

We saw that in the region $A>10$ i.e. $m<M_{\rm Pl}/31.6$, we get a bounded region, with SUGRA point at the boundary. However, when $m$ is close enough to $M_{\rm Pl}$, we get a qualitatively different conclusion.\\ \\ When $A<4$ i.e. $m>M_{\rm Pl}/20$, we get an unbounded region. See for example, Figure  \ref{fig:Unbounded}. The allowed region is unbounded in the $d_2$ direction, and also does not contain the point $(d_1=d_2=0)$.

\subsection{Conclusion}
\begin{figure}[h!]
    \centering
    \includegraphics[width=0.6\linewidth]{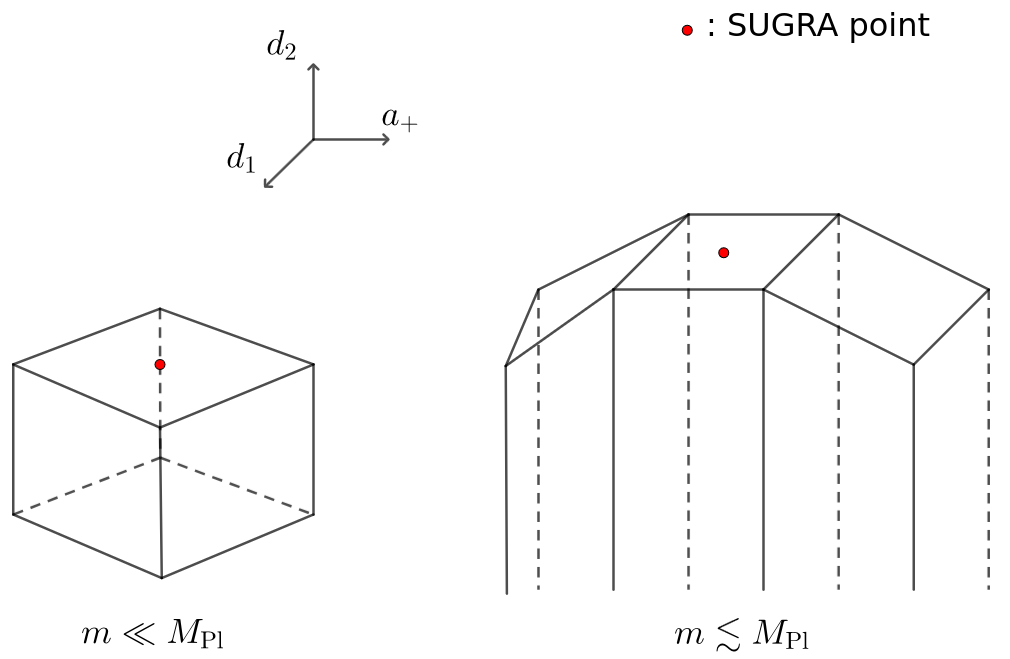}
    \caption{When $m\ll M_{\rm Pl}$, the allowed region is a closed polytope (left) whereas when $m$ is closer to $M_{\rm Pl}$, we get in general an unbounded region in coupling space (right). The Supergravity point in all cases remains on the boundary, although in the latter case, the point $d_1=d_2=0$ may or may not be present in a fixed $a_+=0$ subregion, as depicted by the figure on the right.}
    \label{schematic_summary}
\end{figure}

\begin{figure}[h!]
    \centering
    \includegraphics[width=0.4\linewidth]{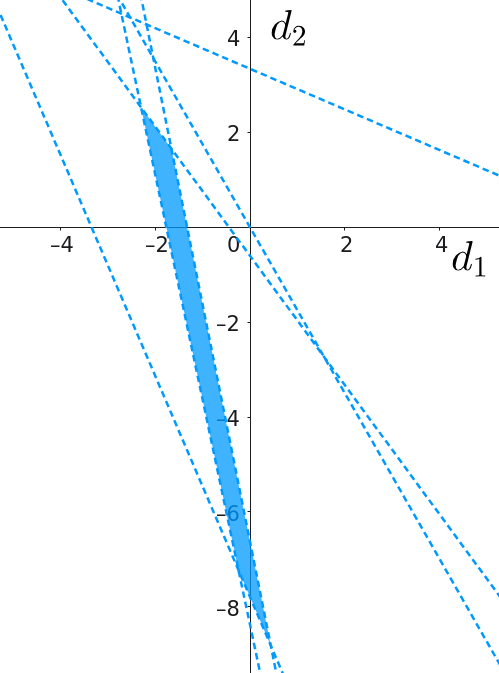}
    \caption{The allowed space of deviations from SUGRA point when $A=5 ~(m\sim M_{\rm Pl}/22.4),~a_+=-1$.}
    \label{fig:A=5}
\end{figure}

This extends the results of \cite{Bellazzini:2025shd}: Even in the small mass limit, there is a small but finite region in EFT parameter space allowed by Unitarity and Analyticity of the longitudinal amplitudes with the Supergravity point living on its boundary. The volume of this region is Planck suppressed : Volume $ \sim m^6/M_{\rm Pl}^6$ and goes to zero in the strict $m\to 0$ limit, agreeing with the results of \cite{Bellazzini:2025shd, Gherghetta:2024tob, Gherghetta:2025tlx}. \\ \\
To reiterate, by ``Supergravity point", we mean $a_+=0, ~a_1=-1/32~~\&~~ a_2=-1/16$; the operator
\begin{equation}
    \frac{a_3}{M_{\rm Pl}^2}(\bar\psi_\nu\gamma^\mu\psi_\mu)^2
\end{equation}
present in the Supergravity Lagrangian \cite{Antoniadis:2022jjy}, vanishes on-shell due to EOM Eq. (\ref{EOM}) and hence contributes only at loop-level and is not constrained by this tree-level analysis. \\ \\
To summarize, we have the following regimes with qualitatively different allowed regions in the $d_1$-$d_2$-$a_+$ coupling space as listed below:\newpage
\begin{itemize}
    \item $\mathbf{A>10}$ ($m<M_{\rm Pl}/31.6$): \\ We get a bounded region (polytope) of volume $\sim m^6/M_{\rm Pl}^6$, with the Supergravity point $(0,0,0)$ on its boundary. See for example, Figures \ref{fig:d1-d2-0} or  \ref{Square}. Schematically, the deviations from Supergravity couplings satisfy
    \begin{equation}\label{d_i bound}
        |d_i|\lesssim \frac{m^2}{M_{\rm Pl}^2}
    \end{equation}
    \item $\mathbf{4<A<10}$ ($M_{\rm Pl}/31.6<m<M_{\rm Pl}/20$): \\The region is unbounded in the $a_+$ direction, but still bounded in (fixed $a_+$) $d_1$-$d_2$ subspace. Depending on the values of $a_+$, the point $d_1=d_2=0$ may not lie within this region. See for example, Figure \ref{fig:A=5}. 
    \item $\mathbf{0.1 <A<4}$ ($M_{\rm Pl}/20<m<M_{\rm Pl}/\sqrt{10}$): \\In this case, the region is unbounded in $d_2$ and $a_+$ directions, and again, depending on the values of $a_+$, the $d_1=d_2=0$ point may not lie within this region. See for example, Figure \ref{fig:Unbounded}.
\end{itemize}
Note that in all these regimes, the Supergravity point $(0,0,0)$ always lies on the boundary of the allowed region in $d_1$-$d_2$-$a_+$ space. \\ \\When the couplings are normalized with respect to the positive coefficient of $s^2$ (say $K$), the space of couplings $\{d_i/K\}$ is bounded even for the case $0.1<A<10$.  If further the gravitational coupling is restored to an arbitrary coupling $g$, we find that positivity allows $g$ to have negative values contrary to what we found in the small mass regime (see discussion after Eq. (\ref{Volume_t})) . Although not strictly an issue, it would be nice to constrain $g>0$ from dispersive arguments in the future, for arbitrary mass regimes.\\ \\
The qualitative structures of the allowed regions are schematically depicted in Figure \ref{schematic_summary}.
\newpage
 \section{Light scalars}\label{Light Scalars}
We saw in the previous section the allowed regions for the contact $\psi^4_\mu$ couplings from dispersive analysis. In this section we perform the dispersive analysis after including a light scalar and pseudo-scalar into the theory to see if the allowed region can be enlarged. The tunings $a_+=0,~a_1=-1/32,~a_2=-1/16$ lead to the cancellation of $E^6$ growth in the $2\to 2$ longitudinal scattering amplitudes, as can be seen from Eqs. (\ref{M++++})-(\ref{M++--}). Since $E^6$ and $E^4$ terms in longitudinal amplitudes are of the type $E^6/m^4M_{\rm Pl}^2$ and $E^4/m^2M_{\rm Pl}^2$ respectively, cancellation of $E^6$ growth raises the (naive) partial wave unitarity cut-off from $(m^2M_{\rm Pl})^{1/3}$ to $\sqrt{mM_{\rm Pl}}$. On top of this, one usually requires the cancellation of $E^4$ growth as well, in order to push the cutoff from $\sqrt{mM_{\rm Pl}}$ to $M_{\rm Pl}$. The simplest way in which this can be achieved is to add a light scalar and a light pseudo scalar in the IR \cite{Gherghetta:2024tob, Antoniadis:2022jjy}.
\begin{equation}
    \mathcal{L}_{\rm int}^{S,P} = \frac{1}{2}g_S \bar\psi_\mu\psi^\mu\phi + \frac{i}{2}g_P \bar\psi_\mu\gamma^5\psi^\mu\Tilde\phi
\end{equation}
We treat the masses of scalars $M_S$ and $M_P$ to be light in the sense that they are smaller than the radius of the contour in Figure \ref{Contour} i.e.
\begin{equation}
    M_S,M_P\ll \sqrt{s_0} <\Lambda
\end{equation}
where $s_0$ is the radius of the contour $\mathcal{C}$ in Figure \ref{Contour}. See footnote \ref{light}. In this case, the $E^4$ contributions in the amplitudes after tuning the couplings $a_+,~a_1,~a_2$ to their SUGRA values, are given below. 
\begin{align}
     \mathcal{M}_{++}^{++} &= -(g_P^2+g_S^2)\frac{s^2}{9m^4}+\frac{s^2}{3m^2M_{\rm Pl}^2} \label{M++++tuned}\\[4pt]
     \mathcal{M}_{+-}^{+-} &= -(g_P^2+g_S^2)\frac{(s+t)^2}{9m^4}+\frac{(s+t)^2}{3m^2M_{\rm Pl}^2}\\[4pt]
     \mathcal{M}_{+-}^{-+} &=+(g_P^2+g_S^2)\frac{t^2}{9m^4}-\frac{t^2}{3m^2M_{\rm Pl}^2}\\[4pt]
     \mathcal{M}_{++}^{--} &= 2(g_S^2-g_P^2)\frac{(t^2-su)}{9m^4} +\frac{16tu}{3M_{\rm Pl}^2s}\label{M++--tuned}
\end{align}
 The second terms are the $E^4$ contributions to graviton $+$ tuned contact amplitude. To cancel the $E^4$ contributions, we must set
\begin{equation}\label{gS,gP point}
    g_S^2=g_P^2=\frac{3m^2}{2M_{\rm Pl}^2}
\end{equation}
These tunings agree with \cite{Gherghetta:2024tob}\footnote{One can show by integration by parts and equations of motion  that the operator considered in \cite{Gherghetta:2024tob}: $\epsilon^{\mu\nu\alpha\beta}(\partial_\mu\Tilde\phi)\, \bar\psi_\nu\gamma_\alpha\psi_\beta$ is equivalent on-shell to $2m\Tilde\phi\,\bar\psi_\mu\gamma_5\psi^\mu$ which we are considering here.}.
We now study the space of deviations from this point, allowed by the dispersive analysis.
\subsection{Positivity bounds on $g_S,~g_P$}
\begin{figure}[h!]
    \centering
    \includegraphics[width=0.5\linewidth]{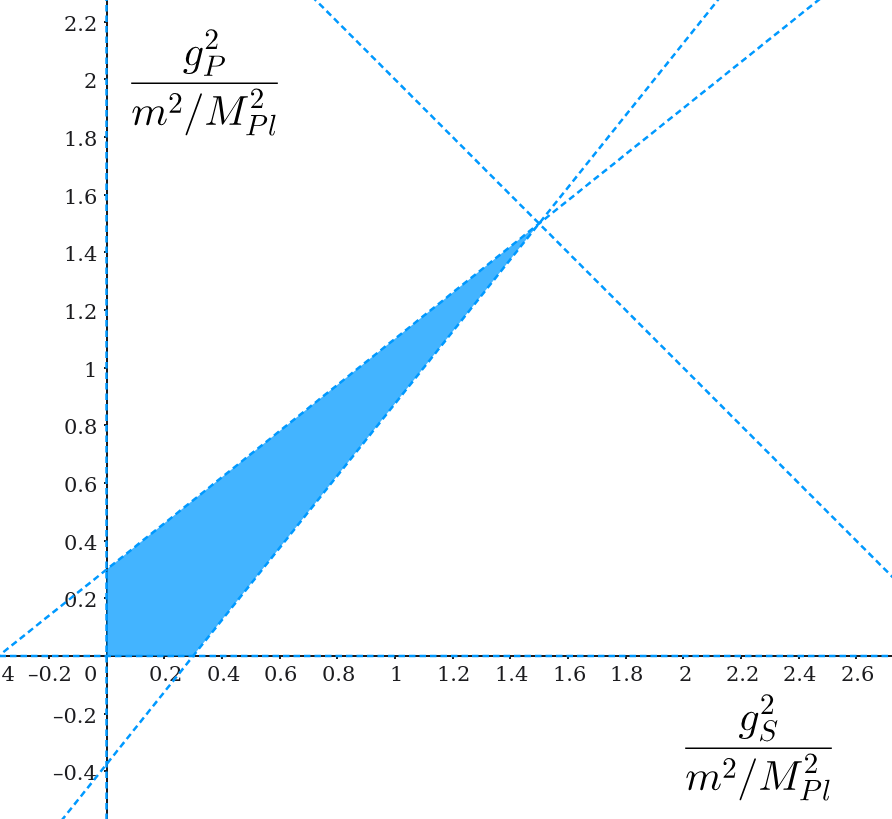}
    \caption{The shaded region shows the allowed values for $g_P^2$ and $g_S^2$ when the contact couplings are tuned to SUGRA values Eq. (\ref{SUGRApoint}). The point Eq. (\ref{gS,gP point}) is at the tip $(1.5,1.5)$ of this region. }
    \label{fig:gP vs gS}
\end{figure}

In this section, we perform the arc analysis on the amplitudes Eqs. (\ref{M++++tuned})-(\ref{M++--tuned}), where the contact couplings are tuned to Supergravity values Eq. (\ref{SUGRApoint}), and derive positivity bounds on $g_S^2$ and $g_P^2$.
The arcs corresponding to these amplitudes are:
\begin{align}
    \mathcal{A}_{++}^{++} &= -\frac{2(g_S^2+g_P^2)}{9m^4}+\frac{2}{3m^2M_{\rm Pl}^2} \\[4pt]
    \mathcal{A}_{++}^{--}&= \frac{4(g_P^2-g_S^2)}{9m^4} \\[4pt]
    \mathcal{A}_{+-}^{-+}&= 0 
\end{align}
Then in the tuned-contact subspace, we get the following bounds on $g_S$ and $g_P$
\begin{align}
    g_S^2 +g_P^2 &\leq \frac{3m^2}{M_{\rm Pl}^2} \\[4pt]
    |g_P^2-g_S^2| &\leq \frac{m^2}{3M_{\rm Pl}^2}-\frac{g_S^2+g_P^2}{9}
\end{align}
These define the region shown in Figure \ref{fig:gP vs gS}.
The area of this region in $g_P^2$-$g_S^2$ space is
\begin{equation}
    \mathrm{Area}=2\frac{m^4}{M_{\rm Pl}^4}
\end{equation}
Thus, we again observe Planck suppressed deviations. Schematically, the relevant couplings $g_S$ and $g_P$ allow for deviations 
\begin{equation}
    \delta g^2\lesssim \frac{m^2}{M_{\rm Pl}^2}
\end{equation}
similar to the contact coupling deviations as we saw in section \ref{Neighbourhood of Supergravity}, Eq. (\ref{d_i bound}).

\subsection{General case}
We now generalize to the case with a graviton, a light scalar and a light pseudo-scalar in the IR, allowing all couplings to deviate from their supergravity values. The interaction Lagrangian is
 \begin{align}
     \mathcal{L}_{\mathrm{int}}&=\frac{1}{M_{\rm Pl}}h_{\mu\nu}T^{\mu\nu}
     +\frac{a_+}{M_{\rm Pl}^2}\left((\bar\psi_\mu\psi^\mu)^2+(\bar\psi_\mu\gamma^5\psi^\mu)^2\right)
     +\frac{ a_1}{M_{\rm Pl}^2}(\bar\psi^\mu \gamma^\rho \psi^\nu)(\bar\psi_\mu \gamma_\rho \psi_\nu)
    \nonumber \\[4pt]
     & +\frac{ a_2}{M_{\rm Pl}^2}(\bar\psi^\rho \gamma^\mu \psi^\nu)(\bar\psi_\rho \gamma_\nu \psi_\mu) 
     +\frac{1}{2}g_S \bar\psi_\mu\psi^\mu\phi 
     + \frac{i}{2}g_P \bar\psi_\mu\gamma^5\psi^\mu\Tilde\phi .
 \end{align}
We parametrize deviations of the scalar couplings as
\begin{equation}
    g_S^2 = \frac{3m^2}{2M_{\rm Pl}^2}(1+d_S),
    \qquad
    g_P^2=\frac{3m^2}{2M_{\rm Pl}^2}(1+d_P).
\end{equation}
In terms of the deviations, the relevant non-forward arcs are
\begin{align}
    \mathcal{A}_{++}^{++} &= -\frac{2}{9m^4M_{\rm Pl}^2}
    \left(4(7d_1+4d_2)t+
    \left(32d_1+32d_2+64a_++\frac{3}{2}(d_S+d_P)\right)m^2\right)
    \\[4pt]
    \mathcal{A}_{++}^{--} &= -\frac{32}{9m^4M_{\rm Pl}^2}
    \left(3a_+t+(6d_1+2d_2)m^2-\frac{3}{16}(d_P-d_S)m^2\right)
    \\[4pt] 
    \mathcal{A}_{+-}^{-+} &=\frac{32}{9m^4M_{\rm Pl}^2}(d_1+d_2)t.
\end{align}
Positivity of the dispersive integrals implies the following linear inequalities:
\begin{align}
    (10+A)d_1+(6+A)d_2+3d_P/8&<(3A-8)a_+, 
    \label{general_bound_1}\\[4pt] 
    (2+A)d_1+(-2+A)d_2-3d_S/8 &> (3A+8)a_+, 
    \\[4pt]
    (-2+A)d_1+(2+A)d_2 +3d_S/8 &< -(3A+8)a_+, 
    \\[4pt]
    (-10+A)d_1+(-6+A)d_2 - 3d_P/8&> -(3A-8)a_+, 
    \\[4pt]
    (16-7A)d_1+4(4-A)d_2+3(d_P+d_S)/4&<-32a_+, 
    \\[4pt]
    7d_1+4d_2&<0.\label{general_bound_6}
\end{align}
 Interestingly, in the subspace $d_S=d_P=0$, i.e. when the scalar couplings are tuned to Polonyi model values, the inequalities Eqs. (\ref{general_bound_1})-(\ref{general_bound_6}) reduce to the bounds when only contact operators are present Eqs. (\ref{ineq1})-(\ref{ineq6}), with $d_1\to a_1$ and $d_2\to a_2$. So, in this subspace, the same conclusions follow as section \ref{Only Contact} but for $d_1,d_2$ instead of $a_1,a_2$ i.e. below a certain mass cut-off ($\sim (0.04)\,M_{\rm Pl}$), the only allowed region is the singular point $(d_1,d_2,a_+)=(0,0,0)$ which is the Supergravity point (as opposed to the finite Planck-suppressed region which was obtained in section \ref{Neighbourhood of Supergravity}). \\ \\
It is interesting that despite minimally coupling to graviton and keeping contact couplings arbitrary, there exists a subspace defined by certain (Polonyi model) values of scalar couplings, where the allowed region (a point) is exactly the same as the case without graviton, just shifted to the Supergravity point from the origin. This can be visualized easily in the subspace $d_S=d_P\equiv -\Delta,~a_+ = 0$. In the small mass approximation, we get
\begin{align}
    -\frac{3\Delta}{8A} &< d_2+ \frac{7}{4}d_1<0 \label{dS=dP_1}\\[4pt]
    -\frac{3\Delta}{16A} &<\frac{d_2+d_1}{2}<\frac{3\Delta}{16A}\label{dS=dP_2}
\end{align}
which is again just the square in Figure  \ref{Square}, with length proportional to $\Delta$. Note that $0<\Delta<1$. Thus, the square shrinks to origin (Supergravity point) when either $\Delta\to 0$ or $m\to 0~(A\to \infty)$. This gives a visual explanation of the observation in the previous paragraph. It is also clear in this subspace, that the allowed region in $d_1$-$d_2$ space is the largest when $\Delta=1\implies g_S=g_P=0$ i.e. when there is no scalar or pseudo-scalar.
\\ \\
In general, in the small mass approximation, the allowed region is approximated by the following inequalities
\begin{align}
    \frac{3d_S}{8A}&<d_1+d_2-3a_+< - \frac{3d_P}{8A} \\[4pt]
    \frac{3d_P}{8A}&<d_1+d_2+3a_+< - \frac{3d_S}{8A} \\[4pt]
    \frac{128a_++3(d_P+d_S)}{4A}&<7d_1+4d_2<0    
\end{align}
From these inequalities, it is also clear that the allowed space of $(d_1,\,d_2,\,a_+)$ is the largest when $d_S=d_P=-1$ i.e. when the scalar and pseudo-scalar are absent.
Qualitatively, we have for the deviations from Supergravity-Polonyi couplings
\begin{equation}
    |\delta a_i| \lesssim \frac{m^2}{M_{\rm Pl}^2}~~~\&~~~\delta g_I^2 \lesssim \frac{m^2}{M_{\rm Pl}^2} ;~~~i=1,2,+,~~I=S,P
\end{equation}
and hence the volume in the 5-dimensional space of deviations $(\delta a_i, \delta g_I^2)$ scales parametrically as
\begin{equation}
    \mathrm{Vol}_5 \sim \left(\frac{m^2}{M_{\rm Pl}^2}\right)^5 
\end{equation}
\section{Partial Wave Unitarity}\label{Partial Wave Unitarity}  
\begin{figure}[h!]
    \centering
    \includegraphics[width=0.5\linewidth]{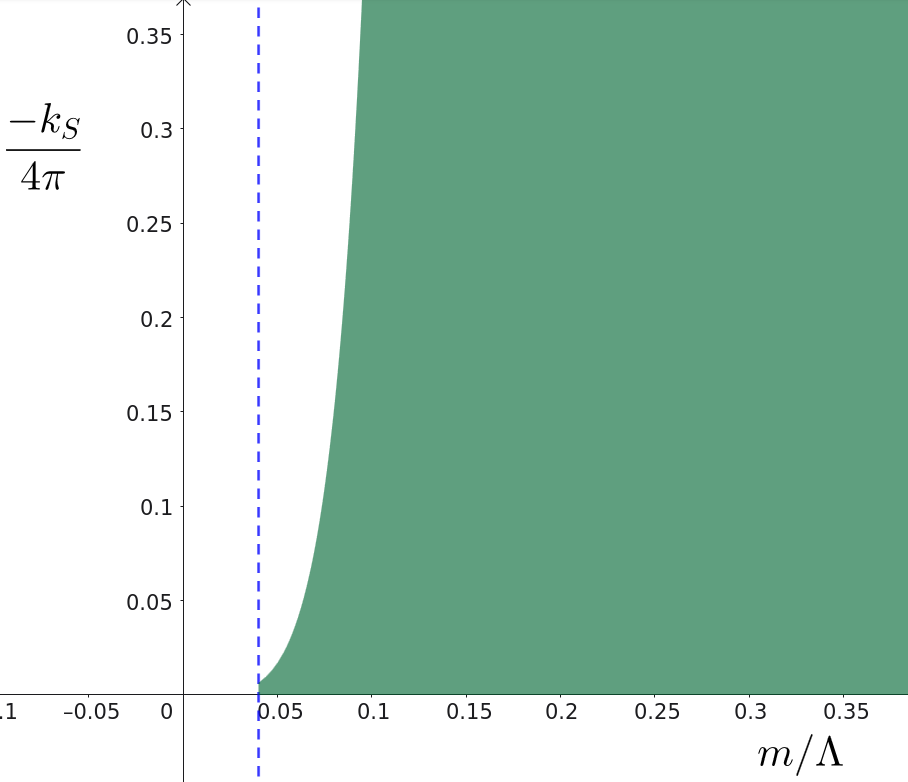}
    \caption{Plot of allowed values of $k_S$ imposing partial wave unitarity up to  $\sqrt{s}\approx (0.3)\Lambda$. We have imposed the dispersive bounds $m/\Lambda>0.04$ (blue line) and $k_S<0$.}
    \label{fig:kS vs m}
\end{figure}

The central assumption/motivation of \cite{Gherghetta:2025tlx} was that partial wave unitarity should hold up to the scale $M_{\rm Pl}$  as discussed in Appendix \ref{on_shell_summary}. However, we argue that demanding partial wave unitarity within the EFT implies bounds on Wilson coefficients, since the unitarity cutoff depends on the couplings. So in this section, we derive bounds on Wilson coefficients from Partial Wave Unitarity and compare the qualitative nature of these bounds with those obtained by dispersive analysis in the previous sections. Performing a partial wave expansion of a $2\to 2$ elastic amplitude 
\begin{equation}
    \mathcal{M}_{\lambda_1\lambda_2}^{\lambda_1\lambda_2} = 16\pi\sum_{j=0}^{\infty}A_j(s) (2j+1) d^j_{\Delta \lambda, \Delta \lambda}\left(\theta\right);~~~\Delta\lambda = \lambda_1-\lambda_2
\end{equation}
the partial wave coefficients $A_j(s)$, assuming perturbative unitarity at tree level, need to satisfy \cite{Schwartz:2014sze, Degrande:2025uil}
\begin{equation}\label{partial_wave_bound}
    |A_j(s)|\leq 1/2
\end{equation}
Using orthogonality relations of Wigner-$d$ matrices, we have
\begin{equation}\label{Partial_Wave}
    A_j(s) = \frac{1}{32\pi }\int_{0}^{\pi} {\rm d}\theta\,\sin\theta~\mathcal{M}(s,\theta)\,d^j_{\Delta \lambda, \Delta \lambda}\left(\theta\right)
\end{equation}
For example, consider the spin-3/2 Fermi interaction.
\begin{equation}
    \mathcal{L}_{\rm int} = \frac{k_S}{\Lambda^2}\left(\bar\psi_\mu \psi^\mu\right)^2
\end{equation}
Corresponding to the amplitude $\mathcal{M}_{++}^{++}$, the $j=0$ partial wave is
\begin{equation}
    A_0(s)|_{++}^{++} = \frac{k_S}{54m^4\pi\Lambda^2}\left(3s^3-22m^2s^2+59m^4s-22m^6\right)
\end{equation}
Unitarity holds up to the scale $s/\Lambda^2$ as long as $|A_0|<1/2$. This gives
\begin{equation}
    \frac{|k_S|}{4\pi}<\frac{1}{2}\,\frac{13.5 \times (m^4/\Lambda^4)}{\left|3\frac{s^3}{\Lambda^6}-22\frac{m^2}{\Lambda^2}\frac{s^2}{\Lambda^4}+59\frac{m^4}{\Lambda^4}\frac{s}{\Lambda^2}-22\frac{m^6}{\Lambda^6}\right|}
\end{equation}
Combining such bounds from all partial waves, the allowed region is shown in Figure \ref{fig:kS vs m} where we have imposed unitarity up to the scale $\sqrt{s}\approx (0.3)\Lambda$ to keep the corrections from higher derivative operators small  enough, as explained in Appendix \ref{Order_of_Corrections}. We are in the regime $m/\Lambda>0.04$ (as required from non-forward positivity, discussed in section \ref{Fermi_EFT}).  We plot $-k_S/4\pi$ instead of $k_S$ since that is what one expects to be the loop expansion parameter, because an $L$-loop diagram contains a factor $(|k_S|/4\pi)^{L+1}$.

\subsection{General case: only contact}
We now perform the analysis on the general case
\begin{equation}\label{L_only_contact_2}
     \mathcal{L}_{\mathrm{int}}=\frac{a_+}{\Lambda^2}\left((\bar\psi_\mu\psi^\mu)^2+(\bar\psi_\mu\gamma^5\psi^\mu)^2\right)+\frac{ a_1}{\Lambda^2}(\bar\psi^\mu \gamma^\rho \psi^\nu)(\bar\psi_\mu \gamma_\rho \psi_\nu)+\frac{ a_2}{\Lambda^2}(\bar\psi^\rho \gamma^\mu \psi^\nu)(\bar\psi_\rho \gamma_\nu \psi_\mu)
 \end{equation}
All relevant partial waves are given in Appendix \ref{Full Amplitudes}.
The region allowed by Partial Wave Unitarity: $|A_j|<1/2$ is bounded as shown below in Figure \ref{fig:Intersection Only Contact}.
\begin{figure}[h!]
    \centering
    \includegraphics[width=0.5\linewidth]{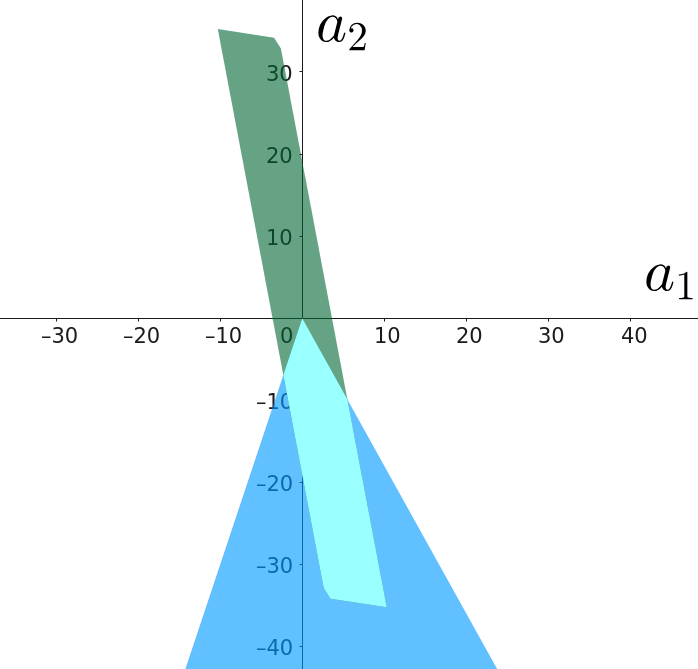}
    \caption{Allowed regions in $a_1$-$a_2$ space from positivity and partial wave unitarity up to the scale $s=(0.1)\, \Lambda^2$, for $a_+=0$ and $m/\Lambda = 0.1$. Green$+$light blue:Partial Wave Unitarity, blue$+$light blue: Positivity. Both bounds are valid upto corrections of around $10\%$.}
    \label{fig:Intersection Only Contact}
\end{figure}
\newpage
\subsection{General case: with the graviton}
\begin{figure}[h!]
    \centering
    \includegraphics[width=0.5\linewidth]{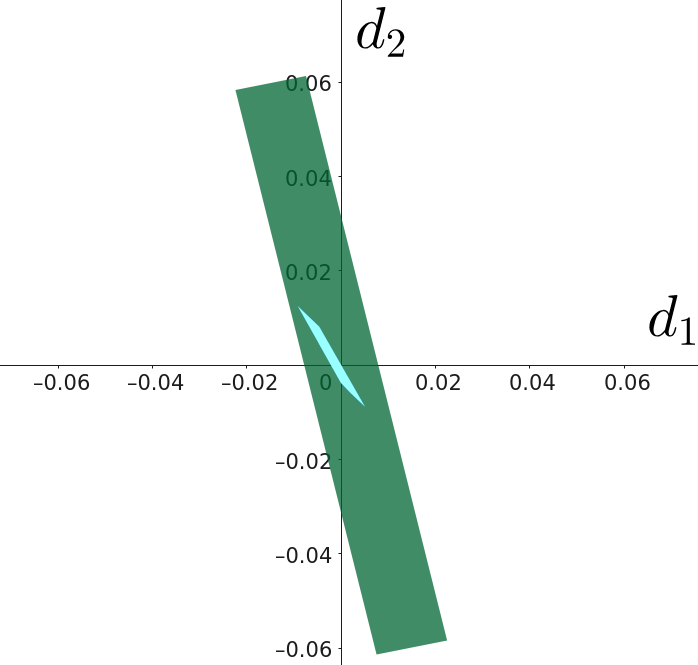}
    \caption{Allowed region in $d_1$-$d_2$ space with $a_+=0,~m=(0.01)M_{\rm Pl},~s=(0.03)M_{\rm Pl}^2$.  Positivity bounds (blue) are accurate upto $1\%$ and the partial wave unitarity bounds (green) are accurate upto $3\%$.}
    \label{fig:SUGRA intersection}
\end{figure}

\begin{figure}[h!]
    \centering
    \includegraphics[width=0.5\linewidth]{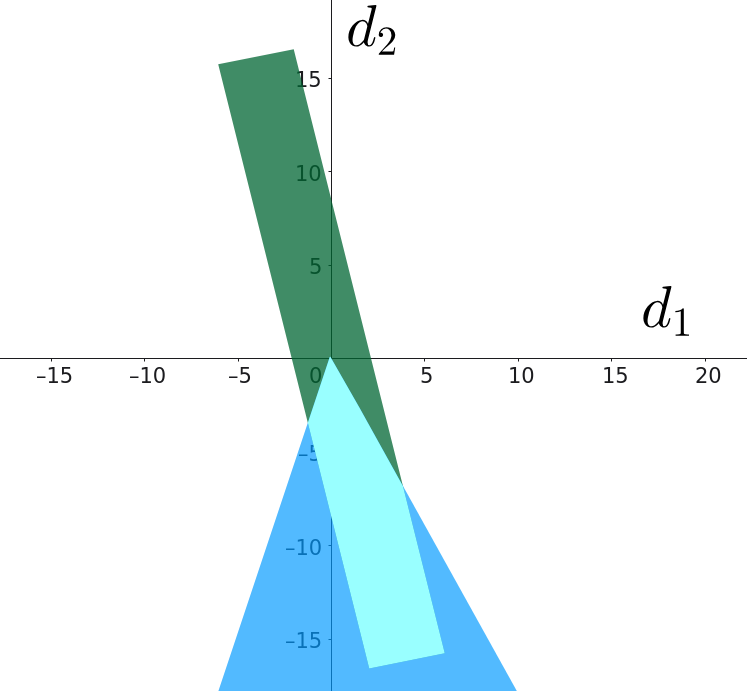}
    \caption{Allowed region in $d_1$-$d_2$ space with $a_+=0,~m=(0.1)M_{\rm Pl},~s=(0.1)M_{\rm Pl}^2$. Green$+$light blue: Partial Wave Unitarity, blue$+$light blue: Positivity. Both bounds are valid upto corrections of around $10\%$.}
    \label{fig:SUGRA intersection 2}
\end{figure}

Next, we perform the partial wave unitarity analysis after adding the minimally coupled graviton  
\begin{equation}\label{L_SUGRA_2}
     \mathcal{L}_{\mathrm{int}}=\frac{1}{M_{\rm Pl}}h_{\mu\nu}T^{\mu\nu}+\frac{a_+}{M_{\rm Pl}^2}\left((\bar\psi_\mu\psi^\mu)^2+(\bar\psi_\mu\gamma^5\psi^\mu)^2\right)+\frac{ a_1}{M_{\rm Pl}^2}(\bar\psi^\mu \gamma^\rho \psi^\nu)(\bar\psi_\mu \gamma_\rho \psi_\nu)+\frac{ a_2}{M_{\rm Pl}^2}(\bar\psi^\rho \gamma^\mu \psi^\nu)(\bar\psi_\rho \gamma_\nu \psi_\mu)
 \end{equation}
 In terms of the deviations from Supergravity couplings
 \begin{equation}\label{devs}
     d_1=a_1+\frac{1}{32},~~d_2=a_2+\frac{1}{16}
 \end{equation}
 the relevant partial waves (upto $\mathcal{O}(E^4)$) are given in Appendix \ref{Full Amplitudes}. The bounds $|A_j|<1/2$ are shown in the plots Figure  \ref{fig:SUGRA intersection} and Figure  \ref{fig:SUGRA intersection 2} for different values of $m$ and $s$. Note that the $s^2/t$ piece in the elastic amplitudes doesn't converge at $\theta=0$. The partial wave coefficients diverge as $\sim E^2\log\left(\sin(\theta_0/2)\right)$ as the lower limit of the integral in Eq. (\ref{Partial_Wave}) $\theta_0\to 0$. We can regulate this divergence by keeping $\theta_0$ such that $E^2|\log(\sin(\theta_0/2))|\ll E^4/m^2$ i.e.
 \begin{equation}\label{IR_log}
     |\log(\sin(\theta_0/2))|\ll \frac{s}{m^2}
 \end{equation}
 For $\theta_0=10^{-5}$, the log factor is $-12$ and for $\theta_0=10^{-10}$, the log factor is $-24$. So in the small mass regime $m^2\ll s$, the $E^2$ piece can be consistently neglected by keeping a small but non-zero lower limit of the integral in Eq. (\ref{Partial_Wave}).
\\
As concluded in section \ref{Neighbourhood of Supergravity}, only the ratios of couplings with respect to $K$ (coefficient of $s^2$ in $\mathcal{M}_{++}^{++}+\mathcal{M}_{+-}^{+-}$)) was bounded in the regime $0.1<A<10$. Partial Wave Unitarity allows us to further bound the absolute value of $K$, since arbitrarily large values of $K$ will result in violating the bounds $|A_j(s)|\lesssim 1$. Thus, the intersection of bounds from Partial Wave Unitarity and Positivity should bound the coupling space completely. This is demonstrated in Figure \ref{fig:SUGRA intersection 2}.
\subsection{Small mass approximation}
We can again get analytical results in the small mass approximation where
\begin{equation}
    m^2\ll s
\end{equation}
In this case, the partial wave unitarity bounds can be approximated by the $s^3/m^4$ pieces of the partial waves. It turns out, only $A_0|_{++}^{++}$ and $A_1|_{+-}^{+-}$ contribute non-trivially to the intersection of bounds from each partial wave.
\begin{align}
    \left|\frac{8(4d_1+d_2)s^3}{432m^4\pi M_{\rm Pl}^2}\right|<\frac{1}{2} ~~\&~~ \left|\frac{16(d_1-5d_2)s^3}{8640m^4\pi M_{\rm Pl}^2}\right|<\frac{1}{2}
\end{align}
Requiring unitarity up to the scale $s/\Lambda^2=1/\Tilde{N}^2$ we get
\begin{equation}
    |4d_1+d_2|<\frac{27}{\pi}\frac{\Tilde{N}^6 m^4}{M_{\rm Pl}^4}~~\&~~|d_1-5d_2|<\frac{270}{\pi}\frac{\Tilde{N}^6 m^4}{M_{\rm Pl}^4}
\end{equation}
The area of the allowed region in $d_1$-$d_2$ space from partial wave unitarity is
\begin{equation}
    A_U = \frac{2430}{7\pi^2}\frac{\Tilde{N}^{12}m^8}{M_{\rm Pl}^8}
\end{equation}
Although this is more Planck suppressed compared to the area obtained by dispersive analysis in the $a_+=0$ subspace Eq. (\ref{positivity_area}), the volume in $d_1$-$d_2$-$a_+$ space is unbounded in the current regime  because $a_+$ is not constrained in the present regime of interest $m\ll \sqrt{s}$. As $m$ approaches $\sqrt{s}$, the analysis becomes inconclusive due to the contribution of the $t$-channel pole. 
\section{Conclusion}
In this paper, we have derived non-forward, tree-level dispersive bounds on effective field theories of massive, Majorana spin-$3/2$ particles, using the dispersive framework of  \cite{Bellazzini:2023nqj_MassiveGravity, Bellazzini:2025shd}.
Our analysis proceeded in three stages: first examining theories with only dimension-six contact interactions, then including a minimally coupled graviton while allowing deviations from the supergravity contact structure, and finally incorporating additional light scalar and pseudo-scalar degrees of freedom.\\\\
We stress again that the dispersive analysis in presence of gravity is valid under the assumption that the $s^2/t$ term, diverging in the forward limit, can be discarded, so that the forward arcs can be defined, as discussed in section \ref{t-pole}. This assumption may be more justified in the regime $m\ll M_{\rm Pl}$ if one considers smearing around $t\sim 0$  \cite{Caron-Huot:2021rmr,Beadle:2024hqg, Beadle:2025cdx, Bellazzini:2025bay} rather than working with exact $t=0$. See the discussion around Eq. (\ref{IR_log}) for a similar argument in context of partial wave unitarity.\\ \\
It is known that, in the small-mass regime, analyticity and positivity enforce the unique $\mathcal{N}=1$ supergravity structure for spin-$3/2$ contact interactions at leading order \cite{Bellazzini:2025shd}. In this work, we study finite-mass deformations around this supergravity solution in greater detail, including subleading effects and geometric properties of the allowed region within the finite-mass dispersive framework of \cite{Bellazzini:2025shd}. We find that, once the gravitino mass is kept non-zero but small compared to $M_{\rm Pl}$, supergravity ceases to be an isolated point in coupling space. Instead, the dispersive constraints admit a bounded neighborhood around the supergravity point, which lies on the boundary of the allowed region.\\\\
In the regime $m \ll M_{\rm Pl}$, the size of this neighbourhood is parametrically suppressed,
\begin{equation}
\mathrm{Vol} \sim \frac{m^{6}}{M_{\rm Pl}^{6}}\, ,
\end{equation}
and shrinks smoothly to zero as $m \to 0$, reproducing the strict massless-limit results. As the mass increases toward the Planck scale, the structure of the allowed region changes qualitatively and unbounded directions in coupling space can emerge.\\\\
We have also investigated whether the inclusion of light scalar and pseudo-scalar fields, motivated by the Polonyi model, can substantially modify this picture. Although such fields can soften the high-energy growth of longitudinal scattering amplitudes and raise the perturbative unitarity cutoff, we find that they do not enlarge the Planck-suppressed neighbourhood around the supergravity point in the small-mass regime. In fact, in the small mass approximation, the allowed (sub)space of couplings $(d_1,\,d_2,\,a_+)$ is the largest in absence of the scalar and pseudo-scalar.  \\\\
Finally, we also derived complementary constraints from the requirement of tree-level partial wave unitarity up to a scale within the EFT. In the small mass regime, we found analytic bounds on two of the three couplings ($d_1$ and $d_2$). They are constrained to lie in a bounded region with no constraints on the third coupling ($a_+$).
\\\\
Taken together, our results provide a quantitative characterization of the finite-mass neighbourhood of supergravity selected by analyticity and unitarity. They show that the strict $m \to 0$ consistency constraints extend to finite mass in a controlled manner, with deviations from the supergravity couplings bounded by powers of $m/M_{\rm Pl}$. 

\appendix
\section{Arc Analysis}\label{ArcAnalysis}
\begin{figure}[h!]
     \centering
     \includegraphics[width=0.5\linewidth]{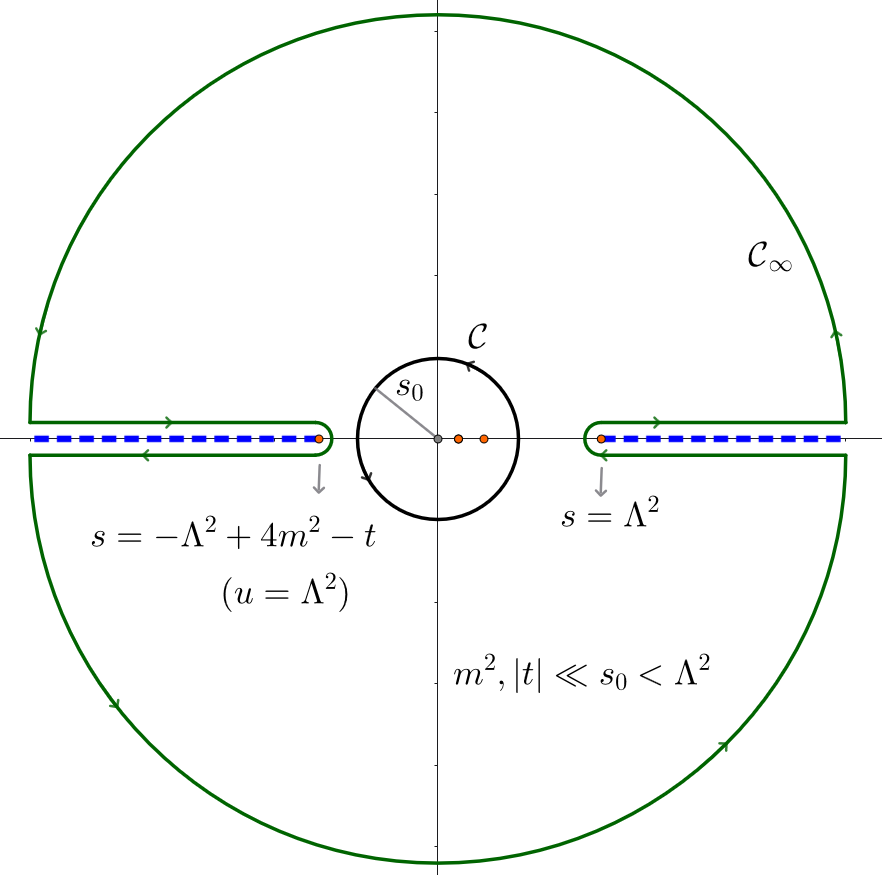}
     \caption{Contours $\mathcal{C}$ and $\mathcal{C}_\infty$ in the complex $s$ plane. The dotted lines represent branch cuts beyond the EFT cut-off $\sim \Lambda^2$. The radius $s_0$ is chosen such that the poles of the integrand in Eq. (\ref{Arc_def}) ($s=4m^2$ and $s=2m^2-t/2$) are within $\mathcal{C}$.}
     \label{Contour}
 \end{figure}

 Here we review the derivation of the non-forward dispersion relations of \cite{Bellazzini:2025shd}, that we have used in this paper. Consider the Arc defined by the integral over the contour $\mathcal{C}$ in the IR and $\mathcal{C}_{\infty}$ in the UV, as shown in Figure \ref{Contour}. The amplitude is assumed to analytic in $s$ for fixed $t\leq 0$ apart from poles and branch cuts on the real $s$ axis\footnote{This was proved by \cite{Bros:1964iho, Bros:1965kbd} up to a finite region $|s|<R(t)$. Then the assumption can be relaxed to requiring $|R(t)|<\Lambda^2$, since in the EFT regime, the singularities are known at tree level.} (see \cite{Martin:1970jsp, Sommer:1970mr, Eden:1971fm} for a review of proven analyticity properties of amplitudes). 
 \\
 The Arc is defined as
 \begin{equation}\label{Arc_def}
    \mathcal{A}_{\lambda_1\lambda_2}^{\lambda_3\lambda_4}(t,n)=\oint_{\mathcal{C}} \frac{ds}{2\pi i}\frac{\mathcal{M}_{\lambda_1\lambda_2}^{\lambda_3\lambda_4}+\mathcal{M}_{\lambda_1\bar\lambda_4}^{\lambda_3\bar\lambda_2}}{(s-2m^2+\frac{t}{2})^{3+n}}=\oint_{\mathcal{C}_{\infty}} \frac{ds}{2\pi i}\frac{\mathcal{M}_{\lambda_1\lambda_2}^{\lambda_3\lambda_4}+\mathcal{M}_{\lambda_1\bar\lambda_4}^{\lambda_3\bar\lambda_2}}{(s-2m^2+\frac{t}{2})^{3+n}}
 \end{equation}
 where the second equality follows because the contour $\mathcal{C}$ can be deformed to $\mathcal{C}_{\infty}$ due to the analyticity assumption. Assuming the Froissart-Martin bound \cite{Froissart:1961ux, Jin:1964zza}: $\lim_{s\to\infty}\mathcal{M}/s^2 = 0$, the circular part of the arc at $\infty$ vanishes. The rest is related to the discontinuities across the branch cuts starting at the EFT cutoff $s=\Lambda^2$ and $s=-t+4m^2-\Lambda^2$ and we get
 \begin{equation}\label{Arc1}
      \mathcal{A}_{\lambda_1\lambda_2}^{\lambda_3\lambda_4}(t,n) = \int_{\Lambda^2}^{\infty}\frac{ds}{\pi}\frac{\mathrm{Im}\mathcal{M}_{\lambda_1\lambda_2}^{\lambda_3\lambda_4}+\mathrm{Im}\mathcal{M}_{\lambda_1\bar\lambda_4}^{\lambda_3\bar\lambda_2} }{(s-2m^2+\frac{t}{2})^{3+n}}
 \end{equation}
 Note that this step of relating the discontinuity across the $s=\Lambda^2$ cut and that across the $u=\Lambda^2$ cut requires the amplitudes in the integrand to be $s$-$u$ symmetric.
 The amplitudes of the type
 \begin{equation}
     \mathcal{M}\equiv \mathcal{M}_{\lambda_1\lambda_2}^{\lambda_3\lambda_4}+\mathcal{M}_{\lambda_1\bar\lambda_4}^{\lambda_3\bar\lambda_2}
 \end{equation}
 are  $s$-$u$ symmetric up to corrections of order $\sqrt{-t}m/s$ in general \cite{TRUEMAN1964322, COHENTANNOUDJI1968239, 10.1143/PTP.45.584} (see \cite{Hebbar:2020ukp} for recent work)
 \begin{equation}
     \mathcal{M}(s,t)=\mathcal{M}(4m^2-s-t,\,t) +\mathcal{O}\left(\frac{\sqrt{-t}m}{s}\right)\sum_{\lambda'_i}C_{\lambda'_1\lambda'_2}^{\lambda'_3\lambda'_4}\mathcal{M}_{\lambda'_1\lambda'_2}^{\lambda'_3\lambda'_4}
 \end{equation}
where $s>\Lambda^2$ inside the UV integral. So  we can consistently ignore these corrections by keeping $|t|, m^2 \ll \Lambda^2$; see Appendix \ref{Order_of_Corrections}. Now using Unitarity i.e. 
 \begin{equation}
     i(\hat T-\hat T^\dagger)=-\hat T^\dagger \hat T
 \end{equation} and the fact that 
\begin{equation*}
     \bra{\Psi}\hat T^\dagger \hat T\ket{\Psi}>0;~~\ket{\Psi}=\ket{p_1,p_2}\pm \ket{p_3,p_4}
 \end{equation*}
 we get
\begin{equation}
    \mathrm{Im}\mathcal{M}_{12}^{12}+\mathrm{Im}\mathcal{M}_{34}^{34} \geq \mp \left(\mathrm{Im}\mathcal{M}_{12}^{34}+\mathrm{Im}\mathcal{M}^{12}_{34}\right) 
\end{equation}
Note that $p_3$ and $p_4$ are chosen in defining $\ket{\Psi}$ such that $p_3+p_4=p_1+p_2$  and the helicity labels are suppressed.
The LHS is positive due to Optical Theorem. So we get
\begin{equation}
    |\mathrm{Im}\mathcal{M}_{12}^{34}+\mathrm{Im}\mathcal{M}_{34}^{12}|\leq \mathrm{Im}\mathcal{M}_{12}^{12}+\mathrm{Im}\mathcal{M}_{34}^{34}
\end{equation}
Using $|A+B|\leq |A|+|B|$ and then the above result in Eq. (\ref{Arc1}), we get
\begin{equation}
    |\mathcal{A}_{\lambda_1\lambda_2}^{\lambda_3\lambda_4}(t,n)|\leq \int_{\Lambda^2}^\infty\frac{ds}{2\pi}\frac{\mathrm{Im}\mathcal{M}_{\lambda_1\lambda_2}^{\lambda_1\lambda_2}+\mathrm{Im}\mathcal{M}_{\lambda_3\lambda_4}^{\lambda_3\lambda_4}+\mathrm{Im}\mathcal{M}_{\lambda_1\bar \lambda_4}^{\lambda_1\bar \lambda_4}+\mathrm{Im}\mathcal{M}_{\lambda_3\bar \lambda_2}^{\lambda_3\bar \lambda_2}\big|_{t=0}}{(s-2m^2+t/2)^{n+3}}
\end{equation}
Then the following combination allows one to relate with the forward arcs
\begin{equation}
      |\mathcal{A}_{\lambda_1\lambda_2}^{\lambda_3\lambda_4}(t,n)|+|\mathcal{A}_{\lambda_1\bar\lambda_2}^{\lambda_3\bar\lambda_4}(t,n)| \leq \frac{1}{2}\left(\mathcal{A}_{\lambda_1\lambda_2}^{\lambda_1\lambda_2}(0,0)+\mathcal{A}_{\lambda_3\lambda_4}^{\lambda_3\lambda_4}(0,0)+\mathcal{A}_{\lambda_1\lambda_4}^{\lambda_1\lambda_4}(0,0)+\mathcal{A}_{\lambda_3\lambda_2}^{\lambda_3\lambda_2}(0,0)\right)
\end{equation}
Then, restricting to longitudinal helicities ($\lambda=\pm 1/2$), and $n=0$, one gets Eq. (\ref{ArcResult}).
\section{Order of Corrections}\label{Order_of_Corrections}
There are expected to be EFT corrections from higher derivative operators in the amplitudes that show up in the bounds derived in this paper from both positivity and partial wave unitarity. Schematically, a longitudinal amplitude with dim-6 and dim-8 EFT operators looks like
\begin{equation}
    \mathcal{M} = C_6\frac{E^6+m^2E^4+...}{m^4\Lambda^2}+ C_8\frac{E^8+m^2E^6+m^4E^4+...}{m^4\Lambda^4}
\end{equation}
and the corresponding arc is (schematically)
\begin{equation}
    \mathcal{A} = C_6\frac{t+m^2}{m^4\Lambda^2}+C_8\frac{t^2 +m^2 t + m^4}{m^4\Lambda^4}
\end{equation}
So in general, the dim-8 Arc is suppressed by $\mathcal{O}(t/\Lambda^2, m^2/\Lambda^2)$ compared to the dim-6 Arc.
\begin{equation}
    \frac{\mathcal{A}(t)}{\mathcal{A}(0)} = \frac{\mathcal{A}_6(t)}{\mathcal{A}_6(0)}\left(1 + \mathcal{O}(m^2/\Lambda^2,~ t/\Lambda^2)\right)
\end{equation}
Thus the bounds also have these corrections, on top of the crossing corrections $\mathcal{O}(\sqrt{-t}m/\Lambda^2)$ discussed in Appendix \ref{ArcAnalysis}.
\begin{equation}
    \frac{|\mathcal{A}_6(t)|}{\mathcal{A}_6(0)}<1 + \mathcal{O}(t/\Lambda^2, m^2/\Lambda^2)+ \mathcal{O}(\sqrt{-t}m/\Lambda^2)
\end{equation}
Similarly, the partial waves also receive corrections from higher derivatives. So the bound Eq. (\ref{partial_wave_bound}) when applied only to the dim-6 operators, is true upto corrections of $\mathcal{O}(s/\Lambda^2)$. \\ \\
So throughout this paper, there are four relevant sources of corrections to the bounds: 
\begin{equation}
    \mathcal{O}\left(\frac{\sqrt{-t} m}{\Lambda^2}\right),~    \mathcal{O}\left(\frac{t}{\Lambda^2}\right),~ \mathcal{O}\left(\frac{m^2}{\Lambda^2}\right),~\mathcal{O}\left(\frac{s}{\Lambda^2}\right)
\end{equation}
The first is from crossing matrix, second and third from higher derivative contribution to the Arcs and the fourth from higher derivative contribution to the partial waves. Thus, to have accuracy in all bounds of around $10\%$, we set  
\begin{equation}
    \frac{\sqrt{-t}}{\Lambda}\Big|_{\rm max}=\frac{1}{N}=0.1 ,~~   \frac{m^2}{\Lambda^2}\Big|_{\rm max} = 0.1~~\&~~ \frac{s}{\Lambda^2}\Big|_{\rm max} = 0.1
\end{equation}
\section{Feynman Rules}\label{Feynman Rules Derivation}
Since it is important to get the overall sign correct for some of the conclusions, we present here a derivation of the Feynman rules used to compute the amplitudes in this paper. 
We first derive the Feynman rules for the operator $(\bar\psi_\alpha \gamma^\mu \psi_\beta)(\bar\psi_\rho \gamma^\nu \psi_\sigma)$. We work with the following conventions, mostly taken from \cite{Peskin:1995ev}. We are evaluating the $12\to 34$ scattering amplitude given by
\begin{align}\label{iM}
    (2\pi)^4\delta(0)i\mathcal{M} &= i\bra{\{p_f\}}\hat T\ket{\{p_i\}} = -i\int d^4x \bra{p_3p_4}\mathcal{H}_{int}\ket{p_1 p_2} \nonumber \\
    &= +i\sqrt{2E_1...2E_4}\int d^4x \bra{0}a_4a_3 (\bar\psi_\alpha \gamma^\mu \psi_\beta)(\bar\psi_\rho \gamma^\nu \psi_\sigma) a_1^\dagger a_2^\dagger\ket{0}
\end{align}
where 
\begin{equation}
    \bra{p_3p_4} = \left(a_3^\dagger a_4^\dagger\ket{0}\right)^\dagger = \bra{0}a_4a_3
\end{equation}
The mode expansion for a Majorana spin-3/2 field is
\begin{equation}\label{3/2ModeExp}
    \psi_\alpha(x) = \int \frac{d^3p}{(2\pi)^3\sqrt{2E_p}}\left(a_p U_\alpha(p)e^{-ipx}+a^\dagger_p V_\alpha (p)e^{+ipx}\right)
\end{equation}
We substitute this in Eq. (\ref{iM}) and in anticipation of momenta going on-shell, cancel the four $\sqrt{2E}$ factors.
\begin{align}
    (2\pi)^4\delta(0)i\mathcal{M} &=  +i\int d^4x \int \frac{d^3\bar k}{(2\pi)^3} \frac{d^3 k}{(2\pi)^3} \frac{d^3\bar p}{(2\pi)^3} \frac{d^3 p}{(2\pi)^3}  \nonumber \\[4pt]
    & \bra{0}a_4a_3 :(a_{\bar k}\bar V_\alpha(\bar k)e^{-i{\bar k}x}+a^\dagger_{\bar k}\bar U_\alpha(\bar k) e^{i{\bar k}x})\gamma^\mu(a_k U_\beta(k) e^{-ikx}+a^\dagger_k V_\beta(k) e^{ikx}) \label{Texpansion} \\[4pt] &\times(a_{\bar p}\bar V_\rho(\bar p) e^{-i{\bar p}x}+a^\dagger_{\bar p}\bar U_\rho(\bar p) e^{i{\bar p}x})\gamma^\nu (a_p U_\sigma(p) e^{-ipx}+a^\dagger_p V_\sigma(p) e^{ipx}) : a_1^\dagger a_2^\dagger\ket{0}\nonumber
\end{align}
 Since only $a^\dagger a^\dagger a a $ type terms will survive, there will be a total of $^4C_2=6$ terms. Rest will annihilate the vacuum. For the Wick contractions, we use
\begin{equation}
    \{a_q, a_r^\dagger \} = (2\pi)^3 \delta^{(3)}(\Vec{q}-\Vec{r})
\end{equation}
For example, the $s$-channel is given by
\begin{align}
    \bra{0}a_4a_3 : a_{\bar k}^\dagger a_{k}^{\dagger} a_{\bar p} a_{p} : a_1^\dagger a_2^\dagger \ket{0}\bar U_{\alpha}(\bar k)\gamma^\mu V_{\beta}(k) \bar V_\rho(\bar p) \gamma^\nu U_\sigma(p) \times e^{i(\bar k + k -\bar p -p)\cdot x} \nonumber \\[4pt]  = 2\times (2\pi)^{3\times 4}\left( \delta_{p,1}\delta_{\bar p,2}\delta_{3,\bar k}\delta_{4, k}-(1\leftrightarrow 2)-(3\leftrightarrow 4)+(1\leftrightarrow 2~\&~3\leftrightarrow 4 )\right)
\end{align}
where the multiplication by 2 takes into account terms where the daggers are put on $a_p$'s instead of $a_k$'s.  \\ \\
So the momenta are put on-shell, the exponentials can be dropped and $\int d^4x$ gives a $(2\pi)^4\delta(0)$. The $(2\pi)^{3\times 4}$ also cancel the four $(2\pi)^3$ in the denominator of Eq. (\ref{Texpansion}). The Dirac-deltas are cancelled by the 3-momentum integrals $\int d^3p$'s. After making these simplifications, we may write schematically
\begin{align}
    i\mathcal{M} &= i \bra{0}a_4a_3 :(a_{\bar k}\bar V_\alpha(\bar k)+a^\dagger_{\bar k}\bar U_\alpha(\bar k) )\gamma^\mu(a_k U_\beta(k) +a^\dagger_k V_\beta(k)) \label{SchematicFeynRule} \\[4pt] &~~~~~~~~~~~~\times(a_{\bar p}\bar V_\rho(\bar p) +a^\dagger_{\bar p}\bar U_\rho(\bar p) )\gamma^\nu (a_p U_\sigma(p) +a^\dagger_p V_\sigma(p)) : a_1^\dagger a_2^\dagger\ket{0}\nonumber
\end{align}
where now, each contraction simply puts the momenta on-shell (without any Dirac-deltas or exponentials). While normal ordering, we have to use
\begin{equation}
    :a_1a_2^\dagger: \,= -a_2^\dagger a_1
\end{equation}
All further signs in different channels are fixed by this.  \\\\
The $s$-channel comes from 2 terms:
\begin{equation}
    \bra{0}a_4a_3 (a_{\bar k}^\dagger a_k^\dagger a_{\bar p} a_p+(k \leftrightarrow p)) a_1^\dagger a_2^\dagger\ket{0}
\end{equation}
The $U,V$ polarization products are determined by Eq. (\ref{SchematicFeynRule}). After all contractions, this gives:
\begin{equation}\label{3/2s}
    (\mathcal{M}_s)_{\alpha\beta\rho\sigma}^{\mu\nu} =  \left(\bar U_\alpha^{(3)}  \gamma^\mu V_\beta^{(4)} - \bar U_\alpha^{(4)} \gamma^\mu V_\beta^{(3)}\right)\left(\bar V_\rho^{(2)} \gamma^\nu U_\sigma^{(1)}-\bar V_\rho^{(1)}\gamma^\nu U_\sigma^{(2)}\right) \times 2
\end{equation}
The other 4 terms 
\begin{equation}
    \bra{0}a_4a_3(-a_{\bar k}^\dagger a_{\bar p}^\dagger a_k a_p+a_{\bar k}^\dagger a_{p}^\dagger a_k a_{\bar p} + (k \leftrightarrow \bar k, p \leftrightarrow \bar p)) a_1^\dagger a_2^\dagger\ket{0}     
\end{equation}
give the $t$ and $u$ channels
\begin{equation}\label{3/2tu}
    (\mathcal{M}_{t,u})_{\alpha\beta\rho\sigma}^{\mu\nu} =
\end{equation}
\begin{align}
     & \Big\{  -(\bar U_\alpha^{(3)} \gamma^\mu U_\beta^{(2)})(\bar U_\rho^{(4)} \gamma^\nu U_\sigma^{(1)}) + (\bar U_\alpha^{(4)} \gamma^\mu U_\beta^{(2)})(\bar U_\rho^{(3)} \gamma^\nu U_\sigma^{(1)})+(\bar U_\alpha^{(3)}\gamma^\mu U_\beta^{(1)})(\bar U_\rho^{(4)} \gamma^\nu U_\sigma^{(2)}) \nonumber\\[4pt]
     &-(\bar U_\alpha^{(4)} \gamma^\mu U_\beta^{(1)})(\bar U_\rho^{(3)} \gamma^\nu U_\sigma^{(2)})\Big\}  +\Big\{(\bar U_\alpha^{(3)} \gamma^\mu U_\beta^{(2)})(\bar V_\rho^{(1)} \gamma^\nu V_\sigma^{(4)}) - (\bar U_\alpha^{(4)} \gamma^\mu U_\beta^{(2)})(\bar V_\rho^{(1)} \gamma^\nu V_\sigma^{(3)})\nonumber \\[4pt]
    &-(\bar U_\alpha^{(3)}\gamma^\mu U_\beta^{(1)})(\bar V_\rho^{(2)} \gamma^\nu V_\sigma^{(4)})+(\bar U_\alpha^{(4)} \gamma^\mu U_\beta^{(1)})(\bar V_\rho^{(2)} \gamma^\nu V_\sigma^{(3)})\Big\} \nonumber \\[4pt]
    &~~~~~~~~~~~~~~~~~~~~~~~~~~~~~~~~~~~~~~~~~~~~~~~~+ (U \leftrightarrow V) \nonumber
\end{align}  
Here, the spinor identities can be shown to hold where along with spin and kinematic labels, the Lorentz indices also get switched, for example
\begin{equation}\label{spinor_identities}
     \bar U_\alpha^{(3)} \gamma^\mu V_\beta^{(4)} =  \bar U_\beta^{(4)} \gamma^\mu V_\alpha^{(3)}~~,~~\bar V_\rho^{(2)} \gamma^\nu V_\sigma^{(4)}=\bar U_\sigma^{(4)} \gamma^\nu U_\rho^{(2)}
\end{equation}
and also
\begin{equation}
    \bar V_\alpha^{(3)} \gamma^\mu V_\beta^{(1)} = \left(U_\alpha^{(3)} \gamma^\mu U_\beta^{(1)}\right)^*
\end{equation}
Using these spinor identities, and the notation $A_{[\alpha} B_{\beta]}\equiv A_\alpha B_\beta-A_{\beta}B_{\alpha}$, we get
\begin{align}
\mathcal{M}_{\alpha\beta\rho\sigma}^{\mu\nu} &= 
      2 \left(\bar U_{[\alpha}^{(3)}  \gamma^\mu V_{\beta]}^{(4)} \right)\left(\bar V_{[\rho}^{(2)} \gamma^\nu U_{\sigma]}^{(1)}\right)  +2\mathrm{Re}\Big[ \left(\bar U_\alpha^{(4)} \gamma^\mu U_\beta^{(2)}\right)\left(\bar U_{[\rho}^{(3)} \gamma^\nu U_{\sigma]}^{(1)}\right) \label{3/2stu} \\[4pt]
     &+\left(\bar U_\alpha^{(3)} \gamma^\mu U_\beta^{(1)}\right)\left(\bar U_{[\rho}^{(4)} \gamma^\nu U_{\sigma]}^{(2)}\right)-\left(\bar U_\alpha^{(4)}\gamma^\mu U_\beta^{(1)}\right)\left(\bar U_{[\rho}^{(3)} \gamma^\nu U_{\sigma]}^{(2)}\right)\nonumber \\[4pt]
     &- \left(\bar U_\alpha^{(3)} \gamma^\mu U_\beta^{(2)}\right)\left(\bar U_{[\rho}^{(4)} \gamma^\nu U_{\sigma]}^{(1)}\right)\Big] \nonumber
\end{align} 
This expression can be further simplified for specific contractions or if one utilizes the symmetry under the exchange $(\mu\alpha\beta)\leftrightarrow(\nu\rho\sigma)$ but will not significantly help with the calculations.
This can be contracted with appropriate metrics to get the amplitudes corresponding to different contact operators in Eq. (\ref{OnlyContactL}).\\ \\
For operators like $(\bar\psi_\mu\psi^\mu)^2$, the Feynman rules can be obtained by making suitable replacements, such as,  replacing $\gamma^\mu,~\gamma^\nu \to 1$ in Eqs. (\ref{3/2s}) and (\ref{3/2tu}). To obtain the simplified version, one can further replace anti-symmetrized `$[]$' indices with symmetrized indices `$\{\}$' in Eq. (\ref{3/2stu}) since Eq. (\ref{spinor_identities}) picks up a minus sign without the $\gamma^\mu$. The Feynman rules for an exchange can be obtained by multiplying with suitable propagators and an additional $-1/2$ coming from $(ig)^2/2!$ in the second order expansion of the $S$-matrix. \newpage
\section{Longitudinal Amplitudes and Partial Wave Coefficients}\label{Full Amplitudes}
In this appendix, we list the $E^6$ and $E^4$ contributions to the scattering amplitudes, corresponding to Eq. (\ref{SUGRAEFT}), which contribute to the Arcs, and the partial wave coefficients relevant to section \ref{Partial Wave Unitarity}.
\subsection*{Amplitudes}
The $E^6$ and $E^4$ contributions to the longitudinal scattering amplitudes corresponding to the Lagrangian Eq. (\ref{SUGRAEFT}) are 
\begin{align}
    \mathcal{M}_{++}^{++} &= \frac{1}{36m^4M_{\rm Pl}^2}\left((1+32a_1)s^3+2stu(3+32a_1+32a_2)\right) \\[4pt] 
    & -\frac{1}{18m^2M_{\rm Pl}^2}\big((3 + 160 a_1 + 64 a_2 + 128 a_+)  s^2 + 2 (17 + 224 a_1 + 160 a_2 + 128 a_+) s t \nonumber\\[4pt]
  &~~~~~~~~~~~~~~~~~~~~+ 2 (23 + 288 a_1 + 224 a_2 + 128 a_+) t^2)\big) \nonumber\\[4pt]
  \mathcal{M}_{+-}^{+-} &= \frac{1}{36m^4M_{\rm Pl}^2}\left((1+32a_1)u^3+2stu(3+32a_1+32a_2)\right) \\[4pt] 
    &+\frac{1}{18m^2M_{\rm Pl}^2s}\big((3 + 32 a_1 - 64 a_2 - 128 a_+) s^3 - (13 + 288 a_1 + 256 a_2 + 
    512 a_+) s^2 t \nonumber\\[4pt] 
    & ~~~~~~~~~~~~~~~~~~- (31 + 544 a_1 + 320 a_2 + 384 a_+) s t^2 - 
 3 (1 + 32 a_1) t^3\big)\nonumber \\[4pt]
  \mathcal{M}_{+-}^{-+} &=- \frac{1}{36m^4M_{\rm Pl}^2}\left((1+32a_1)t^3+2stu(3+32a_1+32a_2)\right) \\[4pt]
    & -\frac{1}{18m^2M_{\rm Pl}^2s}\big(3(1+32a_1)t^3-4t^2s(7+112a_1+80a_2+96a_+) \nonumber\\[4pt]
  &~~~~~~~~~~~~~~~~~~~~- 16 (1+16a_1+8a_2+16a_+) s^2t)\big) \nonumber\\[4pt]
    \mathcal{M}_{++}^{--} &=\frac{16a_+stu}{3m^4M_{\rm Pl}^2} - \frac{1}{9m^2M_{\rm Pl}^2}\big( (5 + 96 a_1 + 32 a_2) s^2 + (19 + 288 a_1 + 160 a_2 + 32 a_+) s t \nonumber\\[4pt]
    &~~~~~~~~~~~~~~~~~~~~~~~~~~~~~~~~~~~~+ (19 + 
    288 a_1 + 160 a_2 + 224 a_+) t^2\big) 
\end{align}\\
Note that the $E^6$ pieces of all amplitudes in this paper satisfy \begin{equation}
    \mathcal{M}_{+-}^{-+}(s,t) = -\mathcal{M}_{++}^{++}(t,s)
\end{equation} 
contrary to the usual assumption $\mathcal{M}_{+-}^{-+}(s,t) = +\mathcal{M}_{++}^{++}(t,s)$. This is fine because the crossing is usually fixed only upto an overall phase, which is fixed by explicit computation (see \cite{Hebbar:2020ukp} for the spin-1/2 case). Regardless, the Arc analysis remains unaffected by this, because only the absolute value $|\mathcal{A}_{+-}^{-+}|$ contributes. Also, this amplitude is not considered in the partial wave unitarity analysis.
\subsection*{Partial wave coefficients}
The partial wave coefficients for the contact terms (corresponding to Eq. (\ref{L_only_contact_2})) are
\begin{align}
     A_0|_{++}^{++} &= \frac{(4 a_1 + a_2) s^3- 
 2 (10 a_1 + 3 a_2 + 8 a_+) m^2 s^2+ (36 a_1 + 5 a_2 + 56 a_+) m^4 s -(17 a_1 + 6 a_2 + 22 a_+) m^6}{54m^4\pi \Lambda^2}\nonumber\\[4pt]
     A_2|_{++}^{++} &= \frac{-( a_1 + a_2) s^3- 
 2 (5 a_1 + 3 a_2 + 4 a_+) m^2 s^2+ (63 a_1 + 43 a_2 + 40 a_+) m^4 s +(17 a_1 + 6 a_2 + 22 a_+) m^6}{270m^4\pi \Lambda^2}\nonumber \\[4pt]
     A_1|_{+-}^{+-} &= \frac{(4 m^2 - s) ((a_1 - 5 a_2) s^2 + 
   2 (-11 a_1 + 5 (a_2 + 2 a_+)) m^2 s + 4 (14 a_1 - 5 a_2) m^4   )}{540m^4\pi\Lambda^2} \nonumber\\[4pt]
     A_2|_{+-}^{+-} &=\frac{(4 m^2 - s) ((a_1 - 5 a_2) s^2+ 
   2 (13 a_1 + 9 (a_2 + 2 a_+)) m^2 s + 4 (5 a_1 + 3 a_2) m^4  )}{540m^4\pi\Lambda^2} \nonumber\\[4pt]
     A_3|_{+-}^{+-} &= \frac{a_1(4m^2-s)(m^2-s)^2}{1890m^4\pi\Lambda^2}\nonumber
\end{align}
and the ones after adding the graviton (corresponding to Eq. (\ref{L_SUGRA_2})), in terms of deviations Eq. (\ref{devs}) are (upto $\mathcal{O}(E^4)$):
 \begin{align}
     A_0|_{++}^{++} &= \frac{8(4d_1+d_2) s^3- 
 (160d_1+48d_2+128a_+ - 9) m^2 s^2+ ...}{432m^4\pi M_{\rm Pl}^2}\nonumber\\[4pt]
     A_2|_{++}^{++} &= -\frac{(s-4m^2)^2\left((d_1+d_2)s^3+2(9d_1+7d_2+4a_+)m^2\right)}{270m^4\pi M_{\rm Pl}^2}\nonumber \\[4pt]
     A_1|_{+-}^{+-} &= \frac{-16(d_1-5d_2)s^3 +(416d_1-480d_2-320a_+ +45)m^2s^2+...}{8640 m^4\pi  M_{\rm Pl}^2} \nonumber\\[4pt]
     A_2|_{+-}^{+-} &=\frac{-16(5d_1+3d_2)s^3-3(32d_1+32d_2+192a_+)m^2s^2+...}{8640m^4\pi  M_{\rm Pl}^2} \nonumber\\[4pt]
     A_3|_{+-}^{+-} &= \frac{-2d_1s^3+12d_1m^2s^2+...}{3780m^4\pi  M_{\rm Pl}^2}\nonumber
\end{align}

\section{Bottom-up on-shell construction of massive spin-$3/2$ theories}\label{on_shell_summary}
In this appendix, we briefly review how the on-shell approach of  \cite{Gherghetta:2024tob,Gherghetta:2025tlx} uniquely fixes the 4-point massive, Majorana spin-3/2 contact interactions to those of $\mathcal{N}=1$ (broken) Supergravity. We translate some of their notations to the ones used in our paper.
\subsection*{Pure spin-$3/2$ contact interactions}
They rule out contact interactions $\psi_\mu^4$  because of the following reason. As shown in Sec~\ref{dim6_EFT}, the longitudinal $2\to 2$ amplitudes behave schematically as
\begin{equation}
\mathcal{M}_{\rm contact}=
\frac{a_i E^6}{m^4 \Lambda^2}
+
\frac{a_i E^4}{m^2 \Lambda^2}
+
\mathcal{O}(E^2/\Lambda^2),
\end{equation}
The leading $E^6$ growth implies a (naive) partial-wave unitarity cutoff
\begin{equation}
\Lambda_{\rm U} \sim (m^2 \Lambda)^{1/3},
\end{equation}
which becomes small in the $m \to 0$ limit and would suggest that unitarity is violated much below the EFT cutoff scale, and hence, they demand a cutoff independent of $m$.\footnote{ However, as explored in detail in section \ref{Partial Wave Unitarity}, the precise statement of partial wave unitarity depends on the precise value of the coupling. The cutoff should actually be 
$ \Lambda_{\rm U} \sim \left(m^2\Lambda/\sqrt{|a_i|}\right)^{1/3}$.
In fact, demanding unitarity implies a bound on the couplings. For some $s\ll\Lambda^2$, 
\begin{equation}
    |a_i|\lesssim \frac{m^4/s^2}{s/\Lambda^2}
\end{equation}
This is made more precise in section \ref{Partial Wave Unitarity}.}\\\\
To achieve this, they add all possible 3-point interactions  in the IR that 
\begin{itemize}
    \item[(i)] have a smooth massless limit and 
    \item[(ii)] have mass dimension $\leq 5$
\end{itemize}
to soften the high energy growth to $\mathcal{O}(E^2/\Lambda^2)$. The corresponding 4-point amplitudes are obtained  using on-shell recursion via the all-line transverse (ALT) shift \cite{Ema:2024vww}, which reconstructs the amplitude from its factorization poles. \\ \\
Constructibility of 4-point amplitudes from 3-point amplitudes is not guaranteed in general. However, \cite{Gherghetta:2024tob} show that in the present context, the 4-point amplitude is constructible provided the Ward identity is imposed on the transverse 4-point amplitudes, in the massless limit.
\subsection*{Spin-0}
Consider first coupling the spin-$3/2$ particle to a scalar and pseudo-scalar. The possible structures of the cubic interactions with mass dimension $\leq 5$ are
\begin{equation}
\frac{g_S}{2}\,\bar{\psi}_\mu \psi^\mu \phi
+
i\,\frac{g_P}{2}
\bar{\psi}_\mu\gamma_5 \psi^\mu\tilde{\phi}
\end{equation}
The corresponding exchange 4-point longitudinal scattering amplitudes in the high energy limit schematically behave as
\begin{equation}
\mathcal{M}_{\phi}
=
-g_S^2 \frac{E^4}{m^4},
\qquad
\mathcal{M}_{\tilde{\phi}}
=
-g_P^2 \frac{E^4}{m^4}.
\end{equation}
There are no contact terms generated under the ALT-shift in these cases. 
\subsection*{Spin-1}
There are three independent  $(\psi_\mu \psi_\mu B_\nu)$ operators with mass dimension $\leq 5$: 
\begin{equation}
    \frac{i g_1}{2\Lambda}\bar\psi_\alpha \gamma^5 \psi_\beta \partial^\alpha B^\beta,~~\frac{g_2}{2\Lambda}\bar\psi_\alpha \psi_\beta \partial^\alpha B^\beta,~~c_B \bar\psi_\alpha \gamma^\mu \gamma^5 \psi^\alpha  B_\mu
\end{equation}
The leading contribution to the 4-point amplitude corresponding to the third operator gives $-E^4$ contribution as shown in \cite{Gherghetta:2025tlx}
\begin{equation}
  \text{e.g.~~} \mathcal{M}_{++}^{++} = -\frac{16c_B^2s^2}{9m^2m_B^2}-\frac{8c_B^2s^2}{9m^4}
\end{equation}
and hence cannot cancel the $-E^4$ contributions from the scalar and pseudo-scalar.\\ \\
We also computed the 4-point amplitude contribution from the first two operators. The leading contributions are:
\begin{equation}
    \mathcal{M}_{++}^{++} = -\frac{(g_1^2+g_2^2) \, s^4 }{36m^4m_B^2\Lambda^2} -\frac{4(g_1^2+g_2^2)\, s^2}{3m_B^2\Lambda^2}-\frac{4g_2^2\, (14s^2t^2+36st^3+26t^4)}{9m_B^2\,s^2\Lambda^2} 
\end{equation}
These operators again give a $-E^4$ contribution, and in addition, a $-E^8$ contribution. Thus, even without assuming a smooth massless limit of the three point amplitudes, one can conclude that these two operators also cannot push the partial wave unitarity upto the scale $M_{\rm Pl}$ by cancelling the $-E^4$ growth.
\subsection*{Spin-2}
There are two independent 3-point interactions (on-shell) with a graviton  with mass-dimension $\leq 5$.
\begin{equation}
   V= h_{\mu\nu}\left[i \xi_1(
     \bar\psi_\rho \gamma^\mu\partial^\nu \psi^\rho-
    \partial^\nu  \bar\psi_\rho \gamma^\mu\psi^\rho)+ i \xi_2(
     \bar\psi^\rho \gamma^\mu\partial_\rho \psi^\nu-
    \partial_\rho \bar\psi^\nu \gamma^\mu\psi^\rho)\right]
\end{equation}
Requiring a smooth massless limit of the spin-3/2 particle (or instead requiring current ($J^\mu=\delta V/\delta \bar\psi_\mu$) conservation in the massless limit, which is a necessary condition for unitarity \cite{Cucchieri:1994tx}) fixes the relative coefficients of these two operators $\xi_2=-2\xi_1$ and leads uniquely to the minimal coupling up to an overall constant $c_h$.
\begin{equation}
c_h \frac{2}{M_{\rm Pl}}\, h_{\mu\nu} T^{\mu\nu}
\end{equation}
where
\begin{equation} 
T^{\mu\nu}= i\left(\bar\psi^\rho \gamma^\mu \partial_\rho \psi^\nu -\partial_\rho \bar\psi^\nu \gamma^\mu \psi^\rho -\frac{1}{2}\bar\psi_\rho \gamma^\mu  {\partial^\nu}\psi^\rho+\frac{1}{2}\partial^\nu\bar\psi_\rho \gamma^\mu\psi^\rho\right)
\label{mincouponshell}
\end{equation}
The four-point amplitude constructed with this interaction via ALT-shift contains both exchange and contact contributions, unlike the spin-0 and spin-1 cases which had only exchange contributions. 
For graviton exchange the amplitude schematically behaves as
\begin{equation}\label{h-exchange}
\mathcal{M}_{(\psi  \psi h)^2}
=
c_h^2  \frac{E^6}{m^4 M_{\rm Pl}^2}
+
\mathcal{O}\left(  \frac{E^4}{m^2 M_{\rm Pl}^2}\right)
+
\mathcal{O}(E^2/M_{\rm Pl}^2).
\end{equation}
In this case, a contact term contribution in the 4-point amplitude is generated under the ALT-shift construction
\begin{align}
 \mathcal{M}_{\psi_\mu^4}= \frac{c_h^2}{2M_{\rm Pl}^2}
    &\Big[(\bar{V}_1 ^\mu \gamma^\nu  U _2^\alpha) (\bar{V}_{3\mu }\gamma_\nu  U_{4\alpha}) -(\bar{V}_1 ^\mu \gamma^\nu  U _2^\alpha) (\bar{V}_{3\alpha }\gamma_\nu  U_{4\mu})+(\bar{V}_1 ^\mu \gamma^\alpha  U _2^\nu) (\bar{V}_{3\alpha }\gamma_\mu  U_{4\nu}) \nonumber\\[4pt]
    &+\,(\bar{V}_1 ^\mu \gamma^\alpha  U _2^\nu )(\bar{V}_{3\mu }\gamma_\nu  U_{4\alpha}) -(\bar{V}_1 ^\mu \gamma^\nu  U _2^\alpha) (\bar{V}_{3\nu }\gamma_\alpha  U_{4\mu})-(\bar{V}_1 ^\alpha \gamma^\nu  U _2^\mu) (\bar{V}_{3\mu }\gamma_\alpha  U_{4\nu}) \nonumber \\[4pt]
    &  -(1 \leftrightarrow 3)-(1 \leftrightarrow 4)\Big]
\end{align} 
which is exactly the $\psi_\mu^4$ gravitino contact term in $\mathcal{N}=1$ Supergravity (corresponding to Eq. (\ref{SUGRAEFT}) with the tunings Eq. (\ref{SUGRApoint})) when one sets $c_h=1/2$. Thus, the Supergravity tunings Eq. (\ref{SUGRApoint}) come out automatically from the ALT-shift construction.\\ \\This contact term gives (for longitudinal scattering)
\begin{equation}
    \mathcal{M}_{\psi_\mu^4} = -c_h^2 \frac{E^6}{m^4M_{\rm Pl}^2} + \mathcal{O}\left(\frac{E^4}{m^2M_{\rm Pl}^2}\right)+
\mathcal{O}(E^2/M_{\rm Pl}^2)
\end{equation}
which cancels the $E^6$ growth of the exchange contribution in Eq. (\ref{h-exchange}) so that
\begin{equation}
    \mathcal{M}_{\rm total}=+c_h^2 \frac{E^4}{m^2M_{\rm Pl}^2}+\mathcal{O}(E^2/M_{\rm Pl}^2)
\end{equation}
and the unitarity cutoff is raised to
\begin{equation}
\Lambda_{\rm U} \sim \sqrt{m M_{\rm Pl}} .
\end{equation}
The remaining $E^4$ contribution has an overall positive sign and thus can be cancelled after combining with spin-0 and spin-1 contributions with specific tunings, and the unitarity cutoff is raised to $M_{\rm Pl}$.  \\ \\
Thus, the following unique features of graviton coupling were responsible for raising the cutoff to Planck scale: 
\begin{itemize}
    \item An $E^6$ contribution in the exchange that can be cancelled by $E^6$ contributions from specific contact interactions, the latter being automatically produced via ALT-shift construction from the 3-point interaction.
     \item A total positive $E^4$ contribution, that could be cancelled by $-E^4$ contributions from spin-0 (and spin-1).
\end{itemize}

\section*{Acknowledgements}
 We thank Farman Ullah for many useful discussions, and collaboration at the initial stage of the project. 
 DG acknowledges support from the Core Research Grant CRG/2023/001448 of the Anusandhan National Research Foundation (ANRF) of the Gov. of India.

\bibliography{biblio}

\end{document}